\documentclass[useAMS,usenatbib]{mn2e}

\usepackage[ansinew]{inputenc}
\usepackage{graphicx} 

\title[A High-Resolution Stellar Library for Evolutionary Population Synthesis]{A High-Resolution Stellar Library for Evolutionary Population Synthesis}
\author[Martins et al.]{Lucimara P. Martins$^{1}$\thanks{E-mail:
martins@stsci.edu}, Rosa M. Gonz\'alez Delgado$^{2}$, Claus Leitherer$^{1}$,\newauthor Miguel Cervi\~no$^{2}$ and Peter Hauschildt$^{3}$\\
$^{1}$Space Telescope Science Institute, 3700 San Martin Dr, Baltimore, MD 21218\\
$^{2}$Instituto de Astrofis\'{\i}ca de Andaluc\'{\i}a (CSIC), Apdo. 3004, 18080 Granada, Spain\\
$^{3}$Hamburger Sternwarte, Gojenbergsweg 112, 21029 Hamburg, Germany }

\begin{document}

\date{Accepted 2004 November. Received 2004 November; in original form 2004 July}

\pagerange{\pageref{firstpage}--\pageref{lastpage}} \pubyear{2005}

\maketitle

\label{firstpage}

\begin{abstract}

We present a library of 1654 high-resolution stellar spectra, with a sampling of 
0.3~\AA\ and covering the wavelength range from 3000 to 7000~\AA. The library was
computed with the latest improvements in stellar atmospheres, incorporating 
non-LTE line-blanketed models for hot, massive ({\it{T$_{\rm{eff}}$}} $\ge$ 27500~K) and line-blanketed
models for cool (3000 K $\le$ {\it{T$_{\rm{eff}}$}}
$\le$ 4500 K) stars. The total coverage of the grid is 3000~K $\le$ {\it{T$_{\rm{eff}}$}} $\le$ 55000~K
and --0.5 $\le$ log {\it{g}} $\le$ 5.5, for four chemical abundance
values: twice solar, solar, half solar and 1/10 solar. 
Evolutionary synthesis models using this library are presented in a 
companion paper. We tested the general behavior of the library by calculating and comparing 
equivalent widths of numerous H and HeI lines, and some of the
commonly used metallic indices. We also compared the library with the empirical libraries 
STELIB and Indo-US.
The full set of the synthetic stellar spectra is available
from our websites (http://www.iaa.csic.es/$\sim$rosa and 
http://www.astro.iag.usp.br/$\sim$lucimara/library.html).

\end{abstract}

\begin{keywords}

stars:atmospheres, stars: evolution
\end{keywords}

\section{Introduction}

Stars are the main energy source in normal galaxies. Their features are detected 
not only in the absorption-line spectra of normal, non-active galaxies, but also 
in starburst and HII galaxies, where emission lines usually dominate. Even in 
active galactic nuclei (AGNs), whose main energy source is gravity rather than 
nucleosynthesis, it has become evident that a significant component of the overall 
energy balance is stellar ionizing radiation (Gonz\'alez Delgado, Heckman, \& 
Leitherer 2001, and references therein). Analyzing the complex spectra of such 
galaxies and disentangling the individual ionizing sources is a major theme of 
contemporary astrophysics and cosmology. Key parameters such as, e.g., metallicity, 
age, or star-formation history allow us to understand how such galaxies form and 
evolve (Kauffmann et al. 2003a).

Over the past years, the quality of observations, both in terms of signal-to-noise 
and spectral resolution underwent a dramatic improvement. Progress has been made 
in the observations of galaxies themselves, but also with respect to data of suitable 
template stars (Le Borgne et al. 2003). The latter aspect is often 
not fully appreciated, but is nevertheless crucial for modeling any galaxy with 
spectral synthesis techniques. Large telescopes with high-resolution spectrographs 
and digital detectors have made it possible to obtain stellar spectra with a 
resolving power of 10$^5$ and sub-percent noise levels for stars down to the 
limit of the Henry Draper catalog. 

This progress has dramatically impacted evolutionary synthesis techniques. 
Such models describe the spectral and chemical evolution of stellar systems 
in an attempt to derive the properties of a stellar population in both nearby 
and distant galaxies (Tinsley 1980). Taking the star-formation history of the 
population (age, initial mass function [IMF], and star formation rate) as a free 
parameter, this technique minimizes the difference between observations and 
models and considers the best-fit solution as the true representation of the 
observations. 

In the past, an analysis of stellar populations has often relied on equivalent 
widths of spectral features, discarding a huge part of the rich information 
present in modern spectra. An example are the widely used Lick indices which 
can be compared to selected features in narrow spectral windows (e.g., 
Worthey et al. 1994). While this technique is still our fundamental key to understanding 
the stellar content of most extragalactic systems (e.g., Trager 2004), global 
synthesis methods fitting the entire spectrum simultaneously are beginning to become 
feasible. An example is the modeling of 10$^5$ nearby galaxies from the Sloan Digital 
Sky Survey by Kauffmann et al. (2003a).

The main challenge of this method is the need of a suitable library of stellar energy 
distributions (SEDs) at high spectral resolution. The available atlases in the 
literature have only intermediate or low 
spectral resolution (e.g., Jacoby, Hunter, \& Christian 1984;  Burstein et al. 1984;
Walborn \& Fitzpatrick 1990; Cananzi, Augarde, \& Lequeux 1993). 
Alternatively, those libraries offering high spectral resolution 
are often limited to a small window of the full spectral range (Cenarro et al. 2001).
 Impressive improvement over previous observational atlases is 
the work of Le Borgne et al. (2003) and Valdes et al. (2004). 
However, the resolution is still only 3 \AA\ or worse,
or these libraries lack completeness in some important stellar evolutionary phases.
The major limitation of all currently available high-resolution empirical libraries 
is the parameter space coverage. A particular concern are chemical abundances. Not 
only are we confined to observations of stars whose heavy-element content is often not 
much different from that of the Sun, but even worse, those stars have the chemical 
evolution of the Galaxy or the Magellanic Cloud imprinted in their spectra. 
Anomalous line strengths due to particular chemical histories have been suggested 
to account for discrepancies between observed and synthetic population spectra 
based on empirical templates (Maraston \& Thomas 2000).

Theoretical stellar spectra do not suffer from this shortcoming. They can be calculated 
for any desired stellar type, luminosity class, chemical abundance, and wavelength 
range. Another important advantage of theoretical models is the knowledge of the 
continuum location: placing the continuum in observed spectra can never be done in 
an unbiased way because the location of the true continuum is unknown. Models predict 
both the line and the continuum spectrum, and the difference between the two allows 
an estimate of the severity of line blanketing. Theoretical libraries 
are already available in the literature, such as, e.g., ATLAS9 (Kurucz 1993b), one
of the most widely used sets of stellar model spectra.
 One of the major drawbacks of this library is its very low
resolution, a shortcoming recently improved by Murphy \& Meiksin (2004). Another
concern with Kurucz models is local thermodynamic equilibrium (LTE),
which is not appropriate for hot stars. Gonz\'alez Delgado \& Leitherer
(1999) avoided this problem by creating a high-resolution synthetic library at solar
chemical abundance using a set of non-LTE models developed by Hubeny and collaborators (Hubeny 1998;
Hubeny, Lanz, \& Jeffrey 1995). However,
computational restrictions limited the spectra to
small spectral ranges around the most important Balmer and HeI lines. 

Motivated by recent progress in atmosphere modeling, we 
have embarked on a project to compute a comprehensive set of stellar atmosphere models 
for implementation in the evolutionary synthesis codes Starburst99\footnote {
http://www.stsci.edu/science/starburst99} (Leitherer et al. 
1999) and sed@\footnote{sed@ is a synthesis code included in the \textbfit{Legacy Tool Project}
 of the \textbfit{Violent Star Formation} European Network more information
 at http://www.iaa.csic.es/$\sim$mcs/sed@}. In the present paper we discuss the choice of the model 
atmospheres and their main characteristics. An accompanying paper (Gonz\'alez Delgado 
et al. 2004, hereafter GD04) provides a parameter study of the stellar population properties computed 
with the new model atmospheres.

This paper is organized as follows: In \S 2 we perform a trade study of the various model 
atmospheres and spectrum synthesis codes in the literature. In \S 3 we discuss the 
generation of the spectral library with particular emphasis on the parameter region 
where the predictions of different models overlap. The major trends of important 
spectral diagnostics with temperature, gravity, and chemical abundance are presented in \S 4. 
In \S 5 we test our theoretical spectra by comparing them to high-quality observations 
of standard stars. Finally, our conclusions are in \S 6.

\section{Theoretical Models}

\subsection{Model atmospheres}

Despite significant improvements over the past years, model atmospheres
are far from perfect in reproducing real stars and still have serious limitations.
Different codes are often optimized for a certain range of parameters ({\it{T$_{\rm{eff}}$}}, log {\it{g}}, etc.), 
and 
not valid for others. Therefore
it is necessary to understand the approximations and applicability limits made by each code
in order to be able to choose between models for different types of stars.
In this section we discuss the codes available in the literature
and justify our preference for this work.

The vast majority of model atmospheres are 1D, time independent and hydrostatic, assume
LTE and treat convection with a rudimentary mixing length. Every one-dimensional
 mixing-length convective model is based on the assumption that the convective structure 
 averages out so that the emergent radiation depends only on a one-dimensional 
 temperature distribution, something that is not always true.
Furthermore, spectral line formation often occurs as a non-equilibrium process: 
under typical atmosphere conditions, radiative rates can dominate over collisional rates, 
and the radiation field departs from the Planck function. Non-LTE line formation is 
therefore neither special nor unusual, while LTE line formation is: LTE is an extreme
 assumption, not a cautious middle-ground.
In general, non-LTE effects become progressively worse for higher temperatures (higher energy) or
very low temperatures (fewer e$^-$), 
lower surface gravities (fewer collisions) and lower [Fe/H] (fewer e$^-$ collisions and stronger UV 
radiation field) (Asplund 2003).

 The microturbulent velocity is another problematic parameter 
that is not calculated self-consistently, except in the Sun. 
Usually it is treated as the parameter that minimizes the scatter among lines of the same 
ion in abundance analyses. It is known that the microturbulent velocity varies with the optical 
depth and that the opacity is strongly depend on microturbulent velocity. Models 
usually assume an average value.

Despite these complications, it is often possible to compute a model that
matches the observed energy distribution and line spectrum of a star. 
However, to obtain the match it is necessary to adjust a number of free parameters: chemical
abundance, effective 
temperature, surface gravity, microturbulent velocity, and the mixing length-to-scale-height-ratio 
in one-dimensional convective treatment.

One of the most widely used libraries is based on Kurucz's (1993b) ATLAS9\footnote{
http://kurucz.harvard.edu/grids.html} model atmospheres.
Kurucz models use the distribution function and line opacity computed by Kurucz 
(1979a,b) from the line data of Kurucz and Peytremann (1975). The latter authors
computed gf-values
for 1.7 million atomic lines for sequences up to nickel. That line list has provided 
the basic data and 
has since been combined with a list of additional lines, corrections, and deletions. 
Kurucz models incorporate line lists for the diatomic molecules 
H$_{2}$, CO, CH, NH, OH, MgH, SiH, CN, C$_{2}$ and TiO. 
In addition to lines between levels, these lists include lines whose wavelengths are 
predicted and are not good enough for detailed spectrum comparisons but are quite 
adequate for statistical opacities. Kurucz recomputed the opacities using these atomic 
and molecular data. Kurucz 
models are not adequate for M stars ({\it{T$_{\rm{eff}}$}} $\le$ 4000 K), because they lack line lists 
or opacities for triatomic molecules (Kurucz 1992).
 
Kurucz models follow from the classical approximations of steady-state, 
homogeneous, LTE, plane-parallel
layers that extend vertically through the 
region where the lines are formed.
In stars later than mid-A, convection can have a significant effect, therefore
models cooler than 9000 K are convective.
The mixing length theory was introduced in ATLAS6 (Kurucz 1979b). This theory is a phenomenological
approach to convection in which it is assumed that one eddy (``bubble'') of a 
given size as a function of local mixing length transports all the convective energy.
One of the shortcomings is the existence of an adjustable parameter $\alpha$, the scale height that a hot bubble rises in the atmosphere before dissipating its heat
to the surrounding gas. The preferred value of $\alpha$ has changed with different ATLAS versions.
In ATLAS9, $\alpha$ is assumed to be 1.25 to fit the energy distribution from the center 
of the Sun. However, the parameter $\alpha$ has to be set at different values to fit different 
types of observations (Steffen \& Ludwig 1999), and no single value works well in all cases.

In ATLAS9, a horizontally averaged opacity and an ``approximate overshooting'' were included. This approximate
overshooting is based on smoothing the convective flux over a certain fraction of the local pressure
scale height at the transition between stable and unstable stratification (Castelli 1999). It 
yields a positive mean convective flux right at the beginning of the stable stratification. 
However, the treatment of convection is still approximate and may be a source
of systematic errors (Gardiner, Kupka, \& Smalley 1999). Gardiner et al. suggested
that the mixing 
length approximation without overshooting works better for stars between 6000 K and 9000 K. 
They also found
that different values of $\alpha$ are necessary for stars in different temperature
ranges ($\alpha$ = 1.25 for 6000 K $\le$ {\it{T$_{\rm{eff}}$}} $\le$ 7000 K, whereas $\alpha$ = 0.5 for
{\it{T$_{\rm{eff}}$}} $\ge$ 7000 K and {\it{T$_{\rm{eff}}$}} $\le$ 6000 K).

Kurucz models also fail where non-LTE and sphericity effects are  important.
Non-LTE effects are much stronger for higher temperatures (O and B stars) and for
low gravities at any temperature (supergiants).
A widely used model atmosphere that takes into account non-LTE
is TLUSTY\footnote{http://tlusty.gsfc.nasa.gov}
 (Hubeny 1988; Hubeny \& Lanz 1995; Lanz \& Hubeny 2003). TLUSTY calculates a 
plane-parallel, horizontally homogeneous model atmosphere 
in radiative and hydrostatic equilibrium. The 
program allows departures from LTE
and metal line blanketing, using the hybrid 
complete linearization and accelerated lambda iteration (CL/ALI)
 method (Hubeny \& Lanz 1995). These models incorporate about 
100000 non-LTE atomic levels and the blanketing effect of millions of Fe and Ni lines.
A total of 8000 lines of the light elements are included, as well as 12 million lines
from Fe III-VI and Ni III--IV. The opacity and radiation field are represented with an 
opacity sampling technique. The convection is treated with the mixing length theory. 
TLUSTY does not account for spherical geometry, which can be important in OB and
Wolf-Rayet stars with strong winds (Aufdenberg et al. 1998, 1999).

A non-LTE code that takes into account stellar winds is WMBasic (Pauldrach et al. 1998, 
Pauldrach, Hoffmann, \&
Lennon 2001). This code calculates expanding atmospheres, taking into account line blocking 
and blanketing. The atomic
data allow a detailed multilevel non-LTE treatment of the metal ions (C to Zn) and
an adequate representation of line blocking and the radiative line acceleration. 
These models also include
EUV and X-ray radiation produced by the cooling zones which originate from the simulation of shock
heated matter.

Phoenix\footnote{http://www.uni-hamburg.de/EN/For/ThA/phoenix/index.html}
(Hauschildt et al. 1996) is a multi-purpose stellar atmosphere code for plane-parallel and spherical models.
The original versions of Phoenix were developed for the modeling of novae and supernova ejecta
(Hauschildt \& Baron 1999 and references therein). Its more recent application to brown
dwarfs is described in detail by Allard \& Hauschildt (1995) and Hauschildt, Allard, \& Baron (1999), and has served
to generate grids of stellar model atmospheres that successfully described low-mass stars in globular
clusters (Baraffe et al. 1995, 1997) and the Galactic disk main sequence (Baraffe et al. 1998).

The equilibrium of Phoenix is solved simultaneously for 40 elements, with usually
two to six ionization stages per element and 600 relevant molecular species for oxygen-rich
ideal gas compositions. The chemistry has been gradually updated with additional 
molecular species since Allard \& Hauschildt
(1995), using the polynomial partition functions of Irwin (1998) and
Sharp \& Huebner (1990). The H$_{3}$$^{+}$ and 
H$_{2}$$^{+}$ ions have been added to the chemical equilibrium and opacity database by using the partition
function of Neale \& Tennyson (1995) and a list of 3 $\times$ 10$^{6}$ transitions 
by Neale, Miller, \& Tennyson
(1996). Van der Waals pressure broadening of the atomic and molecular lines is applied as described
by Schweitzer et al. (1996). Dust is allowed to form, but assumed to dissipate
immediately after formation (Allard et al. 2001). The convective mixing is treated according to the mixing-length theory.
Both atomic and molecular lines are treated with a direct opacity sampling method.

\subsection{The spectrum synthesis codes}

It is important to distinguish between a atmosphere model and a profile
synthesis model. A classical model atmosphere provides 
the run of temperature, gas, electron and radiation pressure, 
convective velocity and flux, and more generally, of all
relevant quantities as a function of some depth variable (geometrical, or optical depth
at some special frequency, or column mass). For comparison with observations, it is 
necessary to calculate the 
synthetic line spectrum from these models with a profile synthesis code. The synthetic spectrum quality
will depend on the quality and adequacy of the atmosphere model, but also
on the details of the line list adopted. Several codes are available in the
literature for this task.

SYNTHE (Kurucz \& Avrett 1981) is a spectrum synthesis program created by Kurucz to 
generate the surface flux for the converged model atmospheres from CD-ROMs 1 and 15 
(Kurucz 1993a,c). The spectrum is computed in 
short intervals, typically at a resolving power of 500,000. Rotationally broadened spectra are computed 
for a number of values of {\it v}sin {\it i}, still at the same resolving power. 

SYNSPEC\footnote{http://tlusty.gsfc.nasa.gov/synspec43/synspec.html} (Hubeny et al. 1995) 
is a general spectrum synthesis program. It requires a model atmosphere as
input. The program reads a general line list and dynamically selects lines which contribute
to the total opacity, based on physical parameters of the actual model atmosphere. SYNSPEC then
solves the radiative transfer equation,
 wavelength by wavelength in a specified spectral 
range, with a specified wavelength resolution. The wavelength points are not equidistant. Instead, 
they are calculated by the program in such a way that there is always a wavelength point at a line
center, and in the midpoint between two neighboring lines. The program then adds a certain number
of points equidistantly spaced between these two, such that the interval between the points does 
not exceed some specified value. This procedure assures that neither any line center nor any
continuum window is  omitted.
The adopted continuum as well as the line opacity is fully specified by the user. In principle,
 the line and continuum opacity sources used in calculating a model stellar atmosphere and in 
 calculating detailed spectrum should be identical, but due to practical limitations, they 
 are usually not.

SPECTRUM\footnote{http://www.phys.appstate.edu/spectrum/spectrum.html}
(Gray \& Corbally 1994) is another routine that computes synthetic spectrum in LTE,
given a stellar atmosphere model. It treats each transition as a pure absorption line.
The code is distributed with an atomic and molecular line list
for the optical spectral region, suitable for computing synthetic spectra
for temperatures between 4500~K and 20000~K. SPECTRUM does not compute adequate spectra
for stars in which a significant part of the line formation occurs in the chromosphere
or in stellar winds. These effects begin to become important in the
Sun and cooler stars. SPECTRUM currently supports the following
diatomic molecules: CH, NH, OH, MgH, SiH, CaH, SiO, C$_{2}$, CN, and TiO. 

There are other codes described in the literature, but they are not suitable for generating
a spectral library containing hundreds of model spectra with full Hertzprung-Russell diagram
(HRD) coverage. 
Routines like MOOG (Sneden 1973) and SME (Valenti \& Piskunov 1996)
 also generate spectral profiles, but their main purpose
is to determine physical parameters of observed spectra by fitting line 
profiles. Owing to this requirements,
they are optimized to generate profiles over very short wavelength 
ranges, and are not
suitable for generating a library as we desire it. 

\subsection{Population synthesis code}

Ideally, one would like to generate a stellar library using model atmospheres
and model spectra that account for all the discussed effects across the full HRD:
non-LTE, line-blanketing, sphericity, expansion, non-radiative heating, convection, etc.
Clearly such an approach is unfeasible -- even if the astrophysical models were available.
The alternative, pragmatic method is to first identify the relevant temperature and gravity ranges
that contribute to the spectrum of a population as predicted by a population synthesis
code and then apply the best physics to those phases that matter the most.

We implemented this library into two stellar population synthesis codes: Starburst99
and sed@. 
The synthesis is done in two steps. First, the code computes the population 
of stars 
as a function of IMF and age. Then the line profiles of the population
are synthesized by adding the luminosity-weighted spectra of individual stars.
By restricting the synthesized spectrum to wavelengths longward of the space-ultraviolet,
we avoid the domain where stellar-wind effects can dominate in hot massive stars. A spectrum
of a young population generated with Starburst99 displays strong stellar-wind lines
of, e.g., CIV and SiIV (Leitherer, Robert, \& Heckman 1995). A code such as WMBasic
would be required for the calculations of these profiles. On the other hand,
a population older than $\sim$5~Myr shows few strong wind lines in the optical, and 
plane-parallel, static models are adequate in the OB-star regime.
Consequently, TLUSTY models are an excellent choice, except for extremely
wind-sensitive lines such as H$\alpha$ or HeII~$\lambda$4686.

At the opposite extreme of the temperature range, line-blanketing becomes 
the dominant issue, and we should strive to apply the best possible physics to
address blanketing. The spherical, fully blanketed Phoenix atmospheres are
the models of choice for cool stars. Late-type stars become noticeable 
in the spectra of a stellar population at wavelengths longward of
$\sim$5000~\AA\ even at young ($\sim$10 Myr) ages, and their influence becomes 
progressively stronger at older ages and longer wavelengths.

While stars at the extreme end of the observed temperature range usually {\it will not}
dominate the total optical light of a stellar population, they {\it will} contribute
with some spectral lines detectable in the spectrum (at least in the populations 
relevant for this paper). Since observed and synthetic galaxy spectra are often
compared with automatic least-square methods, missing even a few of such spectral lines
may have rather significant consequences. Therefore, our rules for selecting
the most appropriate atmospheric and profile models for implementation in our 
evolutionary synthesis codes are as follows:

(i) Line-blanketing must be accounted for at any position in the HRD.

(ii) Non-LTE is most significant at high temperatures.

(iii) Completeness of line lists is important for intermediate-type and paramount 
for cool stars.

This leads us to rely on TLUSTY for O stars, Phoenix for mid-K to M
stars, and Kurucz for the parameter space in between. These choices turn out to be 
most consistent when combined with evolutionary synthesis codes like Starburst99
or sed@.
Details about these codes and the implementation are discussed in another paper
(Gonz\'alez Delgado et al. 2004).

\section{The Library: Technical Details}

Guided by the discussion in the previous section we attempted to create a library as
complete and homogeneous as possible. We decided to concentrate on the optical part of the spectrum,
where most observations are available.
The grid covers the wavelength range 3000 -- 7000~\AA, with 
a sampling of 0.3 \AA, spanning a range of effective temperature from 3000 K --
55000~K, with a variable step from 500 to 2500~K, and surface gravities log {\it{g}} = --0.5 to 5.5, with 
dex steps of 0.25 and 0.5. The library covers chemical abundances of 2 times solar, solar,
half solar and 1/10 solar. The half solar library was constructed with a mix of models with
half solar and one third solar. Models for {\it{T$_{\rm{eff}}$}} 
$\geq$ 8250 K have half solar composition, while the ones for lower {\it{T$_{\rm{eff}}$}}
have one third solar. The He abundances were kept solar (10\% by number density
relative to H) for all models.
The main characteristics of the library are summarized in 
Table~1. This table gives the {\it{T$_{\rm{eff}}$}} and log {\it{g}} coverage and
the atmosphere and profile routines used for the solar abundance models. For the other
abundances, the models are essentially the same, with the following exceptions: 
for models with abundance of 2 times solar, there are no models for {\it{T$_{\rm{eff}}$}} = 3000~K
and log {\it{g}} = --0.5 and 0.0, {\it{T$_{\rm{eff}}$}} = 7500~K and log {\it{g}} = 0.5, 
{\it{T$_{\rm{eff}}$}} = 19000~K and log {\it{g}} = 2.5 and {\it{T$_{\rm{eff}}$}} = 
25000 and log {\it{g}} = 3.5. For half solar abundance, there are no models with {\it{T$_{\rm{eff}}$}} = 3000~K
and log {\it{g}} = --0.5, {\it{T$_{\rm{eff}}$}} = 3500~K and log {\it{g}} = 5.5 and
{\it{T$_{\rm{eff}}$}} = 4000~K and log {\it{g}} =3.5.
For 1/10 solar abundance, there are no models for {\it{T$_{\rm{eff}}$}} = 3000~K
and log {\it{g}} = 3.0 and 3.5. The lack of these library spectra is caused by convergence
problems for the higher temperatures and low gravities. In addition we did not 
generate library spectra at a few  {\it{T$_{\rm{eff}}$}} and log {\it{g}} grid
points where the grid density is sufficiently high to permit equivalent
results.

\begin{table*}
\begin{minipage}{180mm}
\caption{{\it{T$_{\rm{eff}}$}} and log {\it{g}} coverage of the solar abundance grid.}
\begin{center} {\footnotesize
\begin {tabular} {@{}|c|ccccccccccccccc|c|c|c|}
\hline
       & log {\it{g}}&   &    &    &     &   &    &    &    &    &    &    &    & M. Atmosph.& synt.spect.&\\ 
\hline
{\it{T$_{\rm{eff}}$}(K)}&--0.5  &0.0& 0.5& 1.0& 1.5 &2.0& 2.5& 3.0& 3.5& 4.0& 4.5& 5.0& 5.5&          &\\
\hline
  3000 &x     &x  & x  &  x &  x  &  x&  x &   x&   x&  x & x  & x  & x &Phoenix&Phoenix & LTE\\
  3500 &x     &x  & x  &  x &  x  &  x&  x &   x&   x&  x & x  & x  & x &Phoenix&Phoenix & LTE\\
  4000 &x     &x  & x  &  x &  x  &  x&  x &   x&   x&  x & x  & x  & x &Phoenix&Phoenix & LTE\\
  4500 &x     &x  & x  &   x&  x  &  x&  x &   x&   x&  x & x  & x  & x &Phoenix&   Phoenix&LTE \\
  4750 &      &x  & x  &   x&  x  &  x&  x &  x &   x&  x & x  & x  &  &Kurucz&     SPECTRUM  & LTE \\
  5000 &      &x  & x  &   x&  x  &  x&  x &  x &   x&  x & x  & x  &  &Kurucz&     SPECTRUM  & LTE \\
  5250 &      &x  & x  &   x&  x  &  x&  x &  x &   x&  x & x  & x  &  &Kurucz&     SPECTRUM  & LTE \\
  5500 &      &x  & x  &   x&  x  &  x&  x &  x &   x&  x & x  & x  &   &Kurucz&    SPECTRUM  & LTE \\
  5750 &      &x  & x  &   x&  x  &  x&  x &  x &   x&  x & x  & x  &  &Kurucz&     SPECTRUM  & LTE \\
  6000 &      &x  & x  &   x&  x  &  x&  x &  x &   x&  x & x  & x  &   &Kurucz&    SPECTRUM  & LTE \\
  6250 &      &   & x  &   x&  x  &  x&  x &  x &   x&  x & x  & x  &  &Kurucz&     SPECTRUM  & LTE \\
  6500 &      &   &   x&   x&  x  &  x&  x &  x &   x&  x & x  & x  &   &Kurucz&    SPECTRUM  & LTE \\
  6750 &      &   & x  &   x&  x  &  x&  x &  x &   x&  x & x  & x  &  &Kurucz&     SPECTRUM  & LTE \\
  7000 &      &   &  x &   x&  x  &  x&  x &  x &   x&  x & x  & x  &   &Kurucz&    SPECTRUM  & LTE \\
  7250 &      &   & x  &   x&  x  &  x&  x &  x &   x&  x & x  & x  &  &Kurucz&     SPECTRUM  & LTE \\
  7500 &      &   &  x &   x&  x  &  x&  x &  x &   x&  x & x  & x  &   &Kurucz&    SPECTRUM  & LTE\\
  7750 &      &   &    &   x&  x  &  x&  x &  x &   x&  x & x  & x  &  &Kurucz&     SPECTRUM  & LTE \\
  8000 &      &   &    &   x&  x  &  x&  x &  x &   x&  x & x  & x  &    & Kurucz&  SPECTRUM  & LTE \\
  8250 &      &   &    &   x&  x  &  x&  x &  x &   x&  x & x  & x  &    &Kurucz&   SPECTRUM  & LTE \\
  8500 &      &   &    &    &  x  &  x&  x &  x &   x&  x & x  & x  &    & Kurucz&     SYNSPEC & LTE \\
  9000 &      &   &    &    &  x  &  x&  x &  x &   x&  x & x  & x  &    & Kurucz&     SYNSPEC & LTE \\
  9500 &      &   &    &    &     &  x&  x &  x &   x&  x & x  & x  &    & Kurucz&     SYNSPEC & LTE \\
 10000 &      &   &    &    &     &  x&  x &  x &   x&  x & x  & x  &    & Kurucz&     SYNSPEC & LTE \\
 10500 &      &   &    &    &     &  x&  x &  x &   x&  x & x  & x  &    & Kurucz&     SYNSPEC & LTE \\
 11000 &      &   &    &    &     &   &  x &  x &   x&  x & x  & x  &    & Kurucz&     SYNSPEC & LTE \\
 11500 &      &   &    &    &     &   &  x &  x &   x&  x & x  & x  &    & Kurucz&     SYNSPEC & LTE \\
 12000 &      &   &    &    &     &   &  x &  x &   x&  x & x  & x  &    & Kurucz&     SYNSPEC & LTE \\
 12500 &      &   &    &    &     &   &  x &  x &   x&  x & x  & x  &    & Kurucz&     SYNSPEC & LTE \\
 13000 &      &   &    &    &     &   &  x &  x &   x&  x & x  & x  &    & Kurucz&     SYNSPEC & LTE\\
 14000 &      &   &    &    &     &   &  x &  x &   x&  x & x  & x  &     &Kurucz&     SYNSPEC & LTE \\
 15000 &      &   &    &    &     &   &  x &  x &   x&  x & x  & x  &     &Kurucz&     SYNSPEC & LTE \\
 16000 &      &   &    &    &     &   &  x &  x &   x&  x & x  & x  &    & Kurucz&     SYNSPEC & LTE \\
 17000 &      &   &    &    &     &   &  x &  x &   x&  x & x  & x  &    & Kurucz&     SYNSPEC & LTE \\
 18000 &      &   &    &    &     &   &  x &  x &   x&  x & x  & x  &    & Kurucz&     SYNSPEC & LTE \\
 19000 &      &   &    &    &     &   &  x &  x &   x&  x & x  & x  &    & Kurucz&     SYNSPEC & LTE \\
 20000 &      &   &    &    &     &   &    &  x &   x&  x & x  & x  &    & Kurucz&     SYNSPEC & LTE \\
 21000 &      &   &    &    &     &   &    &  x &   x&  x & x  & x  &    & Kurucz&      SYNSPEC& LTE \\
 22000 &      &   &    &    &     &   &    &  x &   x&  x & x  & x  &    & Kurucz &     SYNSPEC& LTE \\
 23000 &      &   &    &    &     &   &    &  x &   x&  x & x  & x  &    & Kurucz&      SYNSPEC& LTE \\
 24000 &      &   &    &    &     &   &    &  x &   x&  x & x  & x  &    & Kurucz &     SYNSPEC& LTE \\
 25000 &      &   &    &    &     &   &    &  x &   x&  x & x  & x  &    & Kurucz&      SYNSPEC& LTE \\
 26000 &      &   &    &    &     &   &    &  x &   x&  x & x  & x  &    & Kurucz&      SYNSPEC& LTE \\
 27000 &      &   &    &    &     &   &    &    &   x&  x & x  & x  &    & Kurucz&      SYNSPEC& LTE \\
 \hline
       &log {\it g} &   &    &    &     &   &    &    &    &    &    &    &     & M. Atmosph.  &synt.spect.&\\
\hline
{\it{T$_{\rm{eff}}$}}(K)&      &   &    &    & 3.00&3.25&3.50&3.75&4.00&4.25&4.50&4.75&&(OSTAR2002)&                    &\\
 \hline
 27500 &      &   &    &    &    x&   x&   x&   x&   x&   x&   x&   x&     & TLUSTY &     SYNSPEC& non-LTE \\
 30000 &      &   &    &    &    x&   x&   x&   x&   x&   x&   x&   x&      &TLUSTY &     SYNSPEC& non-LTE \\
 32500 &      &   &    &    &     &   x&   x&   x&   x&   x&   x&   x&      &TLUSTY &     SYNSPEC& non-LTE \\
 35000 &      &   &    &    &     &   x&   x&   x&   x&   x&   x&   x&     & TLUSTY &     SYNSPEC& non-LTE \\
 37500 &      &   &    &    &     &    &   x&   x&   x&   x&   x&   x&     & TLUSTY &     SYNSPEC& non-LTE \\
 40000 &      &   &    &    &     &    &   x&   x&   x&   x&   x&   x&      &TLUSTY &     SYNSPEC& non-LTE \\
 42500 &      &   &    &    &     &    &    &   x&   x&   x&   x&   x&      &TLUSTY &     SYNSPEC& non-LTE \\
 45000 &      &   &    &    &     &    &    &   x&   x&   x&   x&   x&      &TLUSTY &     SYNSPEC& non-LTE \\
 47500 &      &   &    &    &     &    &    &   x&   x&   x&   x&   x&      &TLUSTY &     SYNSPEC& non-LTE \\
 50000 &      &   &    &    &     &    &    &    &   x&   x&   x&   x&      &TLUSTY &     SYNSPEC& non-LTE \\
 52500 &      &   &    &    &     &    &    &    &   x&   x&   x&   x&      &TLUSTY &     SYNSPEC& non-LTE \\
 55000 &      &   &    &    &     &    &    &    &   x&   x&   x&   x&      &TLUSTY &     SYNSPEC& non-LTE \\
\hline
\end{tabular}}
\end{center}
\end{minipage}
\end{table*}

The grid was constructed as follows: for stars with {\it{T$_{\rm{eff}}$}} $\ge$ 27500~K, we adopted the grid OSTAR2002 
(Lanz \& Hubeny 2003), which was created with TLUSTY model
atmospheres and SYNSPEC spectra. The models are metal line-blanketed, non-LTE, plane-parallel,
and consider hydrostatic atmospheres.  The 
grid covers temperatures ranging from 27500 K to 55000 K, with a 2500 K 
step, and surface gravity in the range 3.0 $\leq$ log 
{\it g} $\leq$ 4.75, with a step of 0.25 dex. Lanz \& Hubeny assumed a microturbulent velocity 
of 10 km s$^{-1}$, which mimics the observed desaturation of 
lines in the photospheric region. WMBasic could be used for these hot stars,
but for the optical wavelength range, TLUSTY has a more detailed and complete metallic line list.

The improvement of non-LTE models over previous models for these stars is significant.
Figure 1 shows a comparison between LTE and non-LTE models for solar chemical abundance.
 From this figure it
is clear that for higher temperatures non-LTE models make a significant difference, as 
already shown by Hauschildt et al. (1999). They argue that LTE models
are adequate for solar-type stars, whereas for cooler and hotter stars the non-LTE effects
become progressively more important.

\begin{figure} 
\includegraphics [width=83mm]{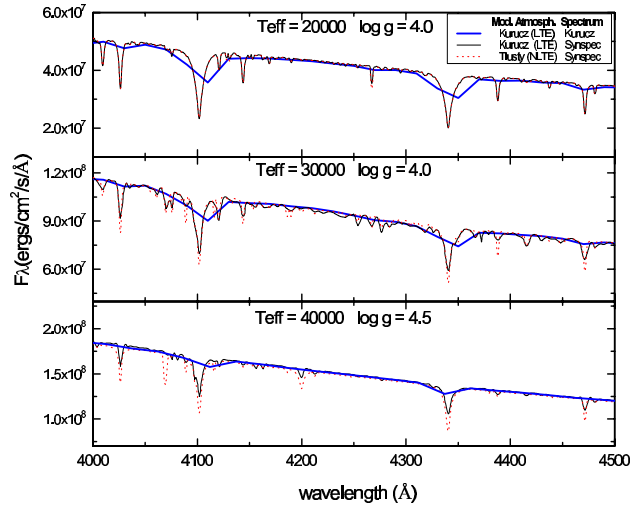}
\caption{Comparison between non-LTE and LTE models for solar abundance. Dashed: Kurucz(LTE)
 at the original published resolution of 20 \AA; dotted: Kurucz (LTE) recomputed with 
 SYNSPEC, at a resolution of 0.3 \AA; solid: TLUSTY (non-LTE) and SYSNSPEC, at a resolution
 of 0.3 \AA. The new non-LTE
models are a dramatic improvement over previous model spectra for {\it{T$_{\rm{eff}}$}} $\ge$ 27500~K.}
\end{figure}

 As LTE effects are still negligible for models cooler than {\it{T$_{\rm{eff}}$}} 
 $\le$ 27000 K, we adopted Kurucz atmosphere 
models for these temperatures. We chose to use the Castelli, Gratton, \& Kurucz (1997) models, which
do not include overshooting. The high-resolution 
spectra were generated using SYNSPEC. We decided to perform the spectrum synthesis
with SYNSPEC instead of SYNTHE (Kurucz \& Avrett 1981) as the former program is better
documented and consequently less prone to user errors (SYNSPEC has a very detailed manual
available online, which is not true for SYNTHE). Besides that, SYNSPEC has some physical
advantadges, as, for example, flexibility in handling atomic data or more
recent hydrogen and helium line broadening tables. We compared these models with previously
generated low resolution Kurucz models, which are known to reliably reproduce the continuum fluxes
for this temperature range (4750 K $\le$ {\it{T$_{\rm{eff}}$}} $\le$ 27000 K). This can be seen in Figure 2,
where our high-resolution models were smoothed with a gaussian filter to mimic
the Kurucz resolution ($\sim$20~\AA). As this
figure shows, our models agree quite well with the Kurucz models down to the temperature
of 7000 K. As demonstrated by Murphy \& Meiksin (2004), the spectra are not expected to be identical, 
since Kurucz models are flux distributions coming directly from the model atmospheres
and not generated by a spectral synthesis.

For {\it{T$_{\rm{eff}}$}} $\le$ 7000~K
molecules start to be
important in the spectra, and we found the latest publicly available version of SYNSPEC
to be inadequate, due to its treatment of the molecular partition functions.
This can
be clearly seen in Fig. 2.
For stars with temperatures between 4750~K and 8250~K, we still used
Kurucz model atmospheres, but now we generated the synthetic profile with the
code SPECTRUM. This code is optimized for this temperature range
and produces reliable results, as shown in Figure 3.
The upper limit was chosen 
based on comparisons between the two models and corresponds to the temperature
where the differences between the two are minimal.

\begin{figure*} 

\includegraphics[width=170mm]{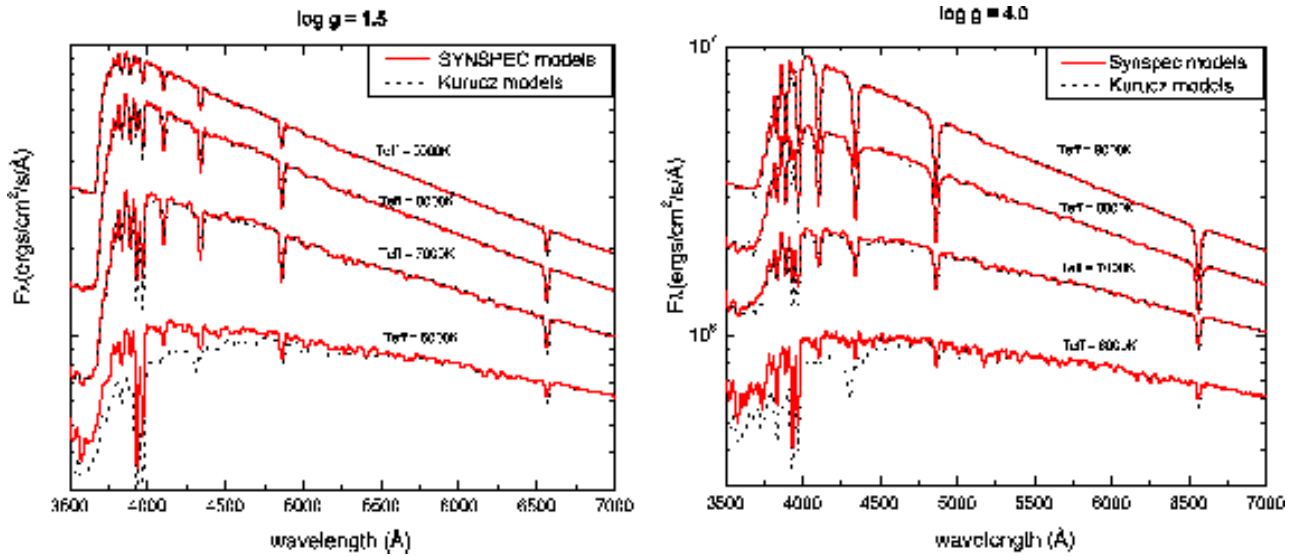}
\caption{Comparison between the SYNSPEC models used in our library and previously
published 
Kurucz models. The SYNSPEC models were smoothed to the resolution of Kurucz models.
Left: log {\it g} =1.5; right: log {\it{g}} = 4.0. SYNSPEC
fails for {\it{T$_{\rm{eff}}$}} $\le$ 7000~K due to inadequate molecular
partition functions. Solar abundance for all
models.}
\end{figure*}

For even lower temperatures ({\it{T$_{\rm{eff}}$}} $\le$ 4500 K)
Kurucz model atmospheres themselves fail. These very cool stars are still 
very hard to reproduce, mainly because molecular features dominate
the spectrum. Phoenix models are a better choice for this temperature range because
they consider
triatomic and larger molecules (676 species in total) and are able to account for spherical geometry.
For models with low gravities (log {\it{g}} $\le$ 3.5),
this can be a dominant effect for the correct calculation of the atmospheric structure
and the synthetic spectrum (Aufdenberg et al. 1998, 1999).
Bertone et al. (2004) compares models from Kurucz's ATLAS9 and Phoenix/NextGen models. 
They argue that ATLAS provides in general a sensibly better fit to observed
spectra of giants and dwarf stars. Their models, however, are different from the ones
used in our library. Bertone et al. used previous-generation Phoenix models, which, for example, 
assume a mixing
length parameter of 1 instead of 2. Two is preferred by hydrodynamic
models. Aufdenberg et al. (1998, 1999) compared interferometry data and the Phoenix
models for M giants with very low gravities, and found a very good agreement.

The publicly available models on the webpage have only intermediate resolution (2~\AA).
Therefore we calculated a new grid with higher resolution.
Our grid of high-resolution spectra was created for
{\it M} = 1 {\it M}$_{\sun}$, a mixing length parameter set to 2.0, spherical geometry,
and constant microturbulent
velocity of 2 km s$^{-1}$. The grid covers {\it{T$_{\rm{eff}}$}} from 3000~K to 4500~K, with a 
500 K step, and --0.5 $\le$
log {\it{g}} $\le$ 5.5. Dust is allowed to form, but assumed to dissipate immediately after formation.
The models are metal line-blanketed. Figure 4 shows a comparison between Kurucz,
Lejeune and Phoenix models. The Lejeune models (Lejeune, Cuisinier, \& Buser 1997)
 were constructed applying a correction function to the original Kurucz models, in order
 to yield synthetic colors matching the empirical color-temperature calibrations derived from
 observations. This correction is only relevant for low-temperature stars ({\it{T$_{\rm{eff}}$}} $<$ 5000 K)
 and is more important for low- than for high-gravity stars.
We emphasize the excellent agreement of the self-consistent Phoenix models
with the empirically adjusted Lejeune
models. 

\begin{figure} 

\includegraphics[width=80mm]{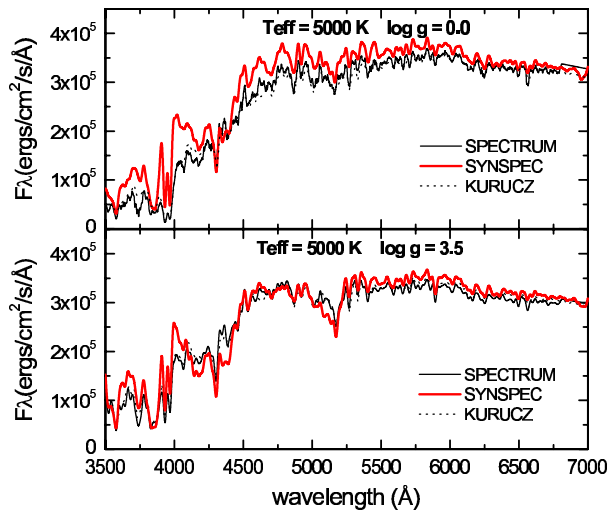}
\caption{Comparison between the Kurucz, SYNSPEC and SPECTRUM models. 
The SYNSPEC  and SPECTRUM models were smoothed to the resolution of Kurucz models. SPECTRUM
generates spectra in excellent agreement with the original Kurucz models.
Solar abundance for all models.}
\end{figure}

\begin{figure} 
\includegraphics[width=83mm]{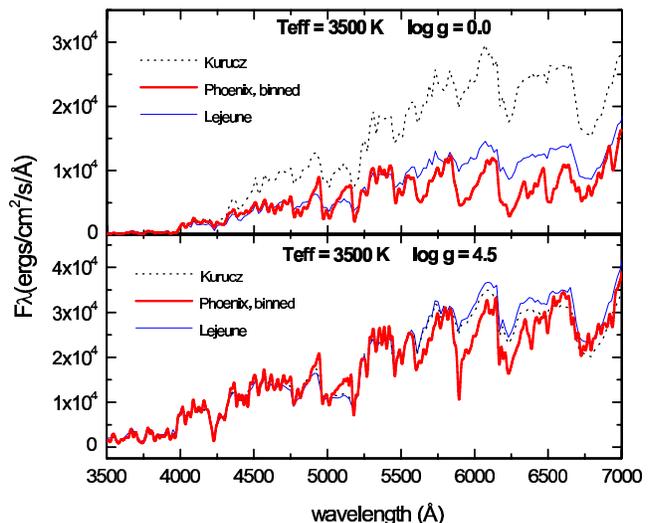}
\caption{Comparison between the Phoenix, Kurucz and Lejeune models.
The Phoenix models were smoothed to the resolution of Kurucz models. The Phoenix models used 
in our library agree with the empirically adjusted Lejeune models. Solar abundance for all models.}
\end{figure}

Models for effective temperatures lower than about 2500~K -- 3000~K need to include the effects
of dust formation and/or dust opacity. This significantly changes the physics of the 
model atmospheres and the formation of the spectrum. From the point of view of stellar
 populations, these stars are not so important because their contribution to the total
luminosity is insignificant. Besides, the evolutionary tracks in Starburst99
and sed@ are not precise 
for this mass range.
Therefore we made no further effort to optimize the library
spectra in this temperature regime.

In all spectra, the maximum distance between two neighboring frequency points for evaluating
the spectrum is 0.01 \AA. The spectra were, however, degraded to a final
resolution of 0.3 \AA, assuming a rotational and instrumentational convolution for each
star. A {\it{T$_{\rm{eff}}$}}-dependent rotational velocity was assumed for all spectra. 
For stars with {\it{T$_{\rm{eff}}$}} $\geq$ 7000~K, we assumed a rotational velocity of 100 km s$^{-1}$.
 For stars with 6000~K $\leq$ {\it{T$_{\rm{eff}}$}} $\leq$ 6500~K, we assumed 50 km s$^{-1}$, and for 
 lower temperatures a value of 10 km s$^{-1}$ was used. These 
 values are typical for stars in these {\it{T$_{\rm{eff}}$}} ranges (de Jager 1980). All the spectra
 are sampled in air wavelengths. Phoenix models were originally generated for vacuum wavelengths, but
 were subsequently transformed to air wavelengths. The final 
 library contains 414, 413, 416 and
 411 spectra for 1/10, half, solar and twice solar chemical composition, respectively 
 (see Table 1).
The grid coverage is illustrated in Figure 5. Isochrones obtained
 with the evolutionary tracks from the Padova group (Girardi et al. 2002)
 have been overplotted. The figure 
 illustrates the homogeneous coverage of this stellar library. The large number of grid points reduces
 the uncertainties of the nearest neighbor assignation assumed in many synthesis
 codes. Except for the Wolf-Rayet phase at the highest {\it{T$_{\rm{eff}}$}} and lowest 
 log {\it g}, our library covers all evolutionary phases.

\begin{figure} 
\includegraphics[angle=270,width=80mm]{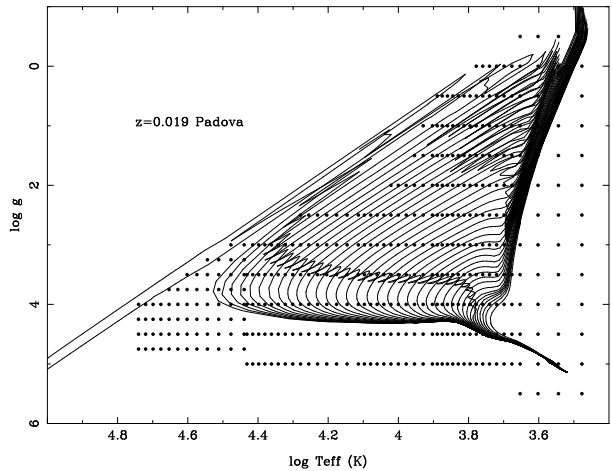}
\caption {Log {\it g} vs. {\it{T$_{\rm{eff}}$}} diagram. Points represent the stellar library grid.
Lines are isochrones of stellar populations that evolve following the Padova
evolutionary tracks with solar chemical abundances. }
\end{figure}

In order to illustrate possible systematic differences in the parts of the
library where we switch the codes used to generate the spectra (stellar atmospheres and/or
numerical method), in Figure 6 we compare spectra at three
effective temperatures (27000 K, 85000 K, and 45000 K), the points where the changes occur. It is
clear from this figure that the transitions are smooth, since the models on this transition points are
very similar. 

\begin{figure*} 
\includegraphics[angle=270,width=160mm]{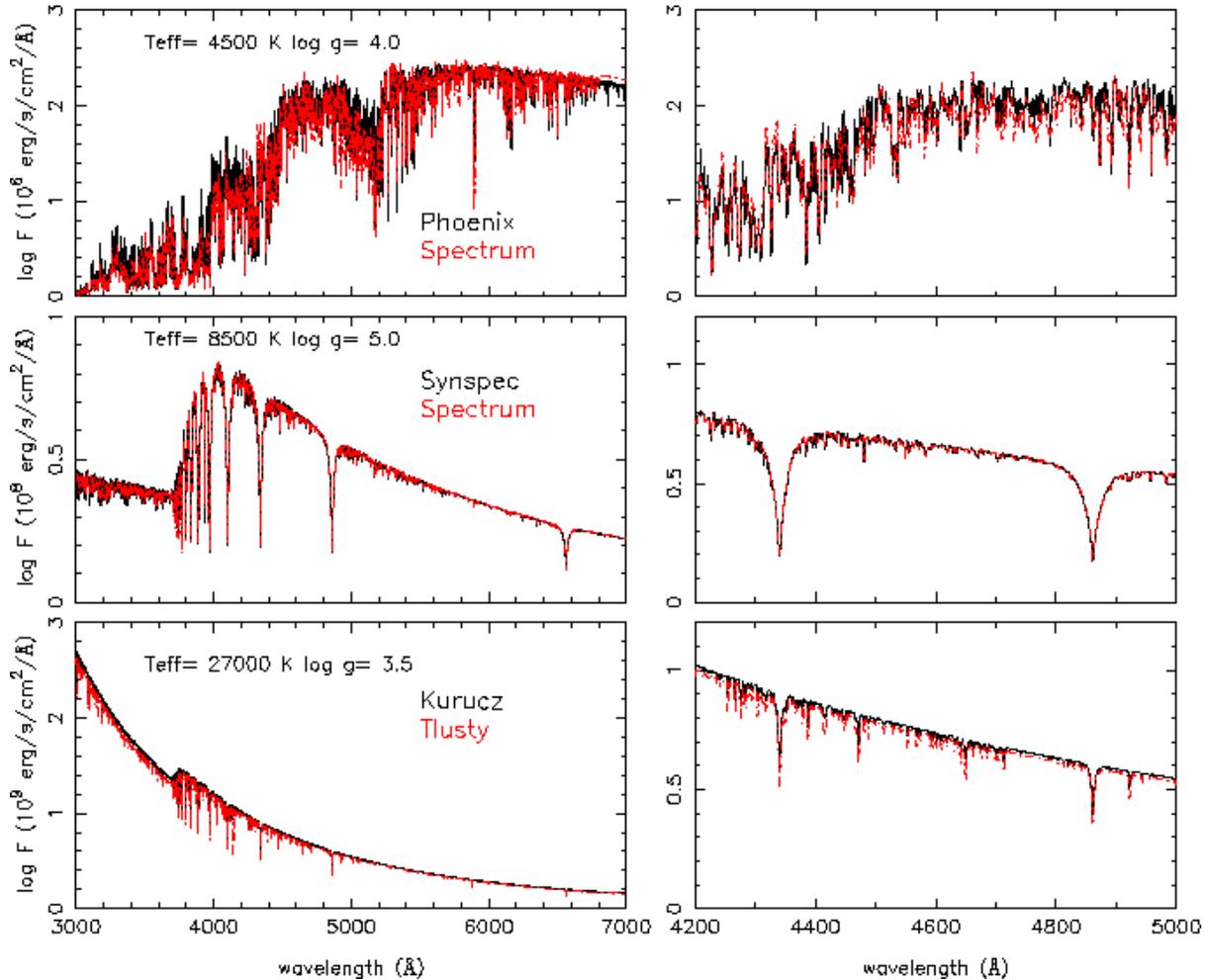}
\caption {Comparison between the spectra obtained with different codes. 
Botton: Teff= 27000 K obtained with 
Kurucz+Synspec (full black line) and Tlusty+Synspec (dotted red line) stellar atmosphere
models. Middle: Teff= 8500 K obtained with Kurucz+Synspec (full black line)
 and Kurucz+Spectrum (dotted red line). Top: Teff= 4500 K obtained with 
Phoenix (full black line) and Kurucz+Spectrum (dotted red line). Solar abundance for
all models.}
\end{figure*}

\section {Properties of the synthesized spectra}

The library is part of the sed@ and Starburst99 codes.
The implementation of the library into the codes is discussed in GD04. The general
behavior of the models is shown on Figures 7 to 9.
Fig.7 shows the variations in the spectra with {\it{T$_{\rm{eff}}$}} for main-sequence
stars with solar abundance. Metallic absorption lines and molecular features
increase with decreasing temperature. The molecular features
dominate the spectra at very low temperatures.
Fig.8 shows the response of the spectra to gravity for a star with 
{\it{T$_{\rm{eff}}$}} = 9000~K and solar abundance. The
width (and equivalent width) of the H lines decreases with gravity. 
Fig.9 illustrates the variation with chemical abundance for a 
main-sequence star with {\it{T$_{\rm{eff}}$}} = 6000~K.  
The blanketing
effect increases from low to high chemical abundances, and from the red to the
blue part of the spectra. 

We also calculated the equivalent widths for the most important H and He lines
in these synthetic spectra. We tested different spectral
windows of 20~\AA, 30~\AA\ and 60~\AA\ for the hydrogen lines, 
in order to measure the contribution of weaker lines
adjacent to the spectral index. The 
comparison between these windows can be seen in Figure 10,
where we have plotted the predicted H$\delta$ equivalent widths
for the three window sizes. Fig.10 suggests a rather significant
influence of the window size due to the inclusion of numerous,
weak metallic absorption lines. For the following discussion, all
measurements refer to the 30~\AA\ window, which we adopted as
the default for all Balmer-line measurements. Table 2 summarizes
the window definitions.
A small discontinuity at
log {\it{T$_{\rm{eff}}$}} = 4.4 ({\it{T$_{\rm{eff}}$}} = 27500 K) is visible in Fig.10. This is
the temperature where we switch to the OSTAR2002
grid, meaning that this is where we switch from non-LTE to LTE. 
Besides that, the models in this grid were calculated with a very detailed and extended line list,
accounting for the blanketing effects of millions of Fe and Ni lines. Therefore
weak lines appear in these
models, falling into the line windows we used and increasing the equivalent widths.
This effect is present in Figures 11 -- 14 as well.

We measured the H and HeI equivalent widths in two ways:
(a) using the theoretical continuum, calculated by each code used, that would 
give the equivalent widths of the lines in a strict theoretical definition; 
(b) using a pseudo-continuum, which was determined by 
fitting a first-order polynomial to the windows defined in Table 2. This simulates 
the values that we could measure observationally. The equivalent widths for these 
lines were calculated
only for stars with {\it{T$_{\rm{eff}}$}} $\ge$ 5000~K, since at lower temperature
the H lines become very weak,
and the measurement would be contaminated by the surrounding metallic lines. 
For example, at {\it{T$_{\rm{eff}}$}} = 4500 K and log {\it{g}} = 4.0, the H$\gamma$ line center is only
5\% below the continuum value.

 \begin{figure} 
\includegraphics[scale=0.5]{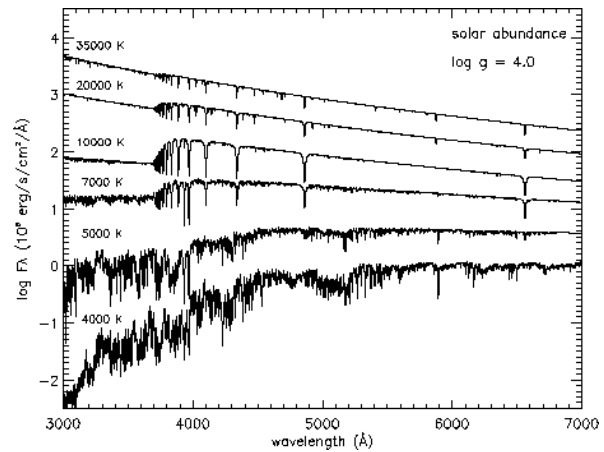}
\caption{Dependence of representative spectra on {\it{T$_{\rm{eff}}$}}.}
\end{figure} 

\begin{figure} 
\includegraphics[scale=0.5]{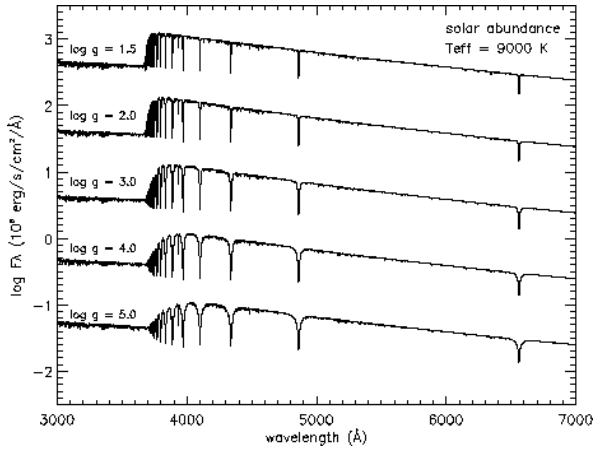}
\caption{Dependence of representative spectra on log {\it{g}}. The spectra were shifted
by a constant value for clarity.}
\end{figure} 

 \begin{figure} 
\includegraphics[scale=0.5]{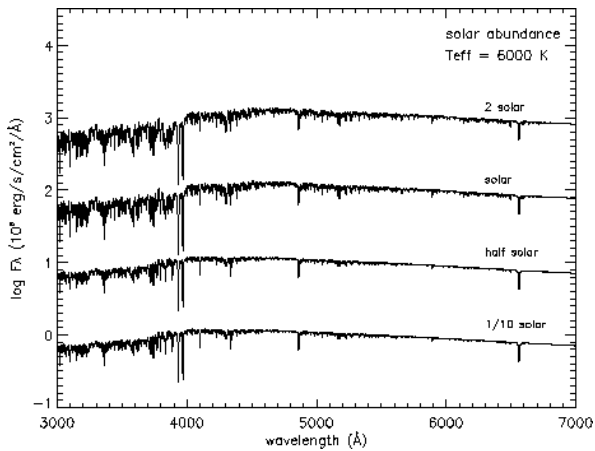}
\caption{Dependence of representative spectra on the chemical composition. The spectra were shifted
by a constant value for clarity.}
\end{figure}

\begin{table*}
\begin{minipage}{110mm}
\caption{Line and continuum windows used for the equivalent width measurements.}
\begin {tabular} {@{}|l|ccc}
\hline
          &  Blue Continuum & Line &  Red Continuum \\
\hline
H$\delta$ & 4012.1 -- 4019.9 & 4087.1 -- 4117.1 & 4157.9 -- 4169.0\\     
H$\gamma$ & 4262.0 -- 4270.1 & 4325.0 -- 4355.0 & 4445.0 -- 4453.1\\
H$\beta$  & 4769.9 -- 4781.9 & 4847.0 -- 4877.0 & 4942.1 -- 4954.1\\
H$\alpha$ & 6506.0 -- 6514.1 & 6548.0 -- 6578.0 & 6611.9 -- 6620.0\\
HeI $\lambda$4026\footnote{This line is actually the sum of HeI
$\lambda$4026 and HeII $\lambda$4025.6. The latter becomes dominant for very early O-type stars.} & 4012.1 -- 4019.9 &  4019.9 -- 4031.0 & 4157.9 -- 4169.0\\
HeI $\lambda$4471 & 4445.0 -- 4453.1 &  4463.9 -- 4478.0 & 4493.0 -- 4503.0\\
HeI $\lambda$5876 & 5834.9 -- 5845.1 &  5870.9 -- 5879.9 & 5903.9 -- 5912.0\\
\hline
\end{tabular}
\end{minipage}
\end{table*}

\begin{figure} 
\includegraphics[scale=0.5]{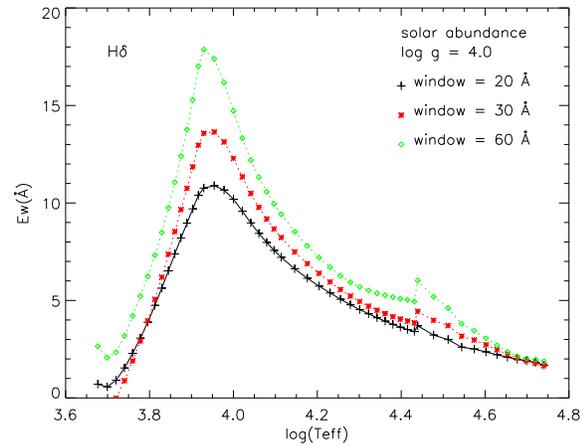}
\caption{Comparison between the H$\delta$ equivalent width measured with different apertures.
The measurements are for log {\it{g}} = 4 and solar abundance. The equivalent width was calculated as
the integrated flux under a pseudo-continuum defined after fitting a linear function to the continuum
windows defined in Table 2. The 30 \AA\ line window is defined in Table 2. The boundaries for the 20~\AA\
and 60~\AA\ windows are 4092 -- 4112~\AA\ and 4070 -- 4130~\AA, respectively.}
\end{figure}

The equivalent widths of H$\delta$, H$\gamma$, H$\beta$ and H$\alpha$ as a function
of {\it{T$_{\rm{eff}}$}} for main-sequence stars (log {\it{g}} = 4.0) at solar abundance
are shown for the theoretical and pseudo-continuum in the left and right panels
of Figure 11, respectively. The 
equivalent widths calculated with either method agree
quite well for stars with {\it{T$_{\rm{eff}}$}} $\ge$ 7000 K. For lower temperatures,
molecular lines and bands are very strong, and the placement of the pseudo-continuum
does not agree with the theoretical continuum anymore. While the pseudo-continuum is affected
by absorption, the real continuum is not, which causes an increase in the equivalent
widths measured with this method. 

Figure 12 shows the dependence of the 
H$\gamma$ equivalent width on gravity. The equivalent width of this figure was 
calculated using the pseudo-continuum. H$\gamma$ (and all the other Balmer lines)
are rather sensitive to gravity and temperature, making them an efficient tool for the 
determination of the fundamental
stellar parameters.

\begin{figure*} 
\includegraphics[angle=90,scale=0.7]{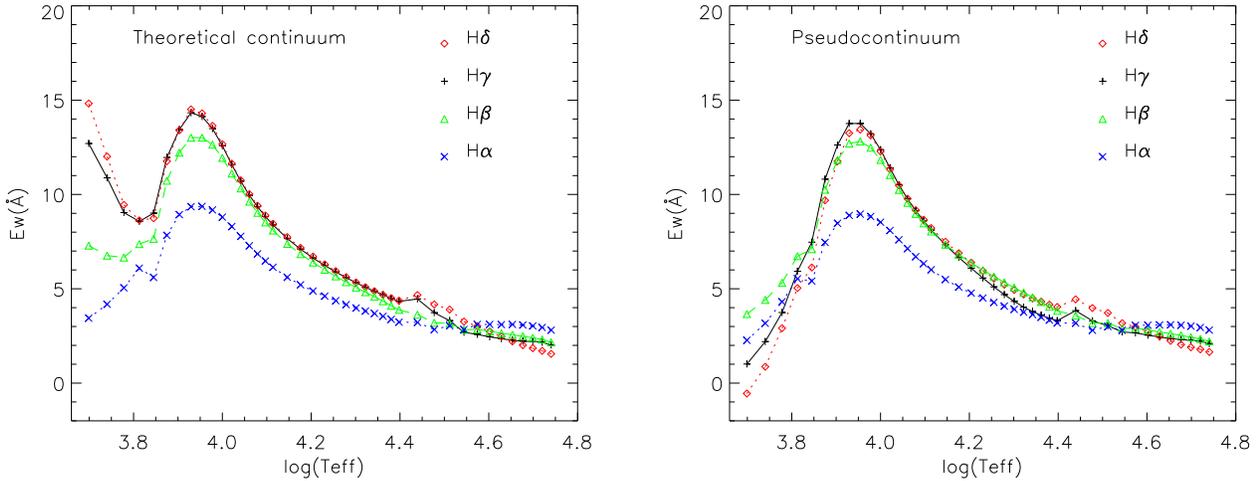}
\caption{Equivalent width of the Balmer lines as a function of the effective 
temperature for log {\it{g}} = 4.0 and solar abundance.
The equivalent widths were measured in windows of 30~\AA. The left panel shows 
the equivalent width calculated as the integrated flux under
the real continuum. The right panel shows the equivalent width calculated
as the integrated flux under a pseudo-continuum defined after fitting a linear function to the continuum
windows defined in Table 2.}
\end{figure*}

\begin{figure} 
\includegraphics[scale=0.5]{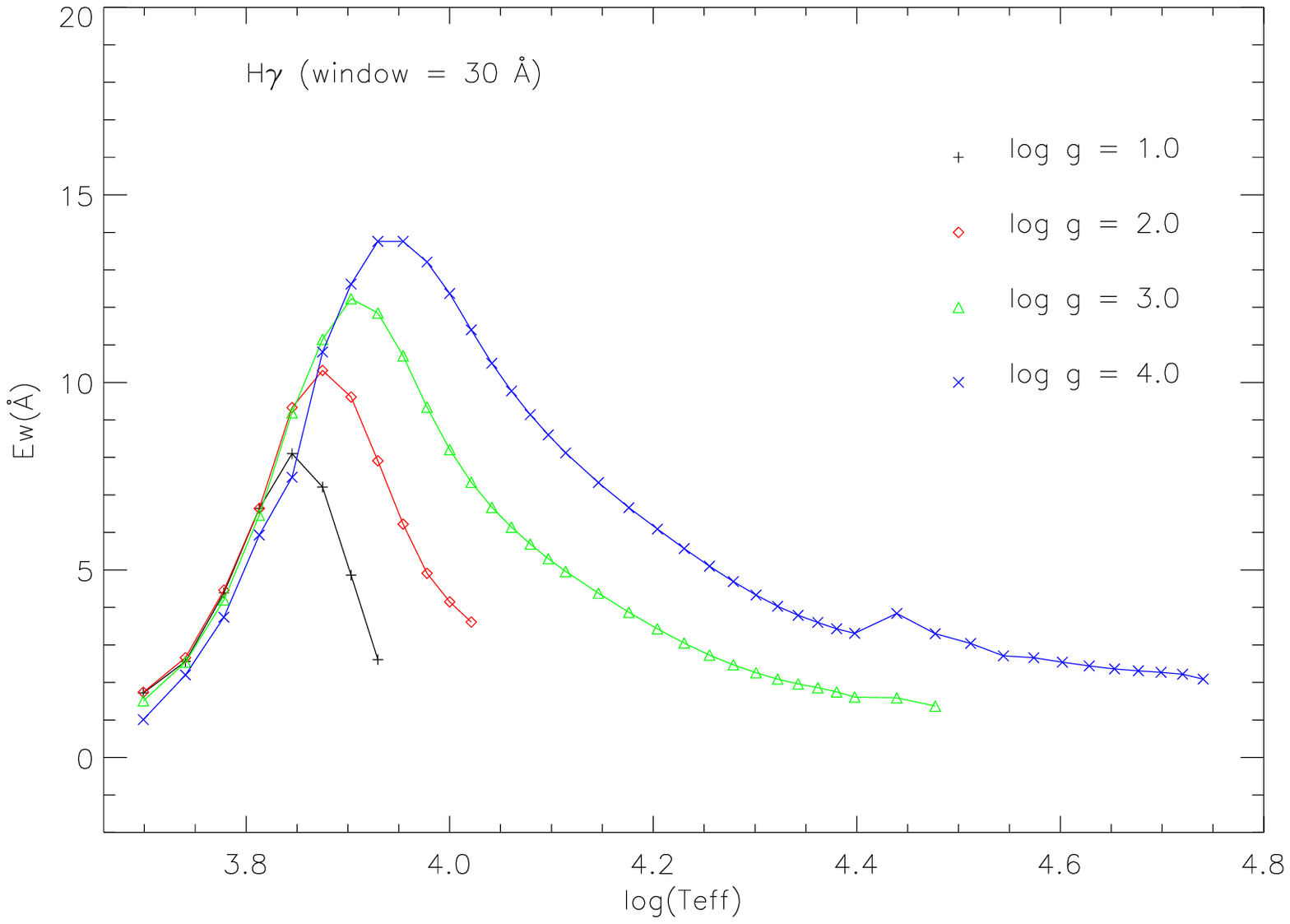}
\caption{Equivalent width of H$\gamma$ as a function of the effective temperature.
Each curve corresponds to a different value of the gravity. The chemical abundance is solar and the 
width of the window used to compute the equivalent widths is 30 \AA.}
\end{figure}

Figure 13 shows the equivalent width of selected HeI lines 
as a function of {\it{T$_{\rm{eff}}$}} for main-sequence stars. The definition of the 
continuum is the same as before.
The dependence on gravity is in Figure 14. The equivalent width was
calculated using the pseudo-continuum.
There is an increase of the equivalent width at {\it{T$_{\rm{eff}}$}} $<$ 10000 K 
which is not due to an increase in HeI absorption,
but to the increase of metallic lines that fall into the
windows used for the equivalent widths measured. The equivalent 
width of these lines is insignificant for these temperatures.
Therefore we truncated the graphs at {\it{T$_{\rm{eff}}$}}
= 10000~K
in Fig.13 and 14.

As expected, the equivalent widths of the H and HeI lines are not affected by chemical abundance changes.

\begin{figure*} 
\includegraphics[angle=90,scale=0.7]{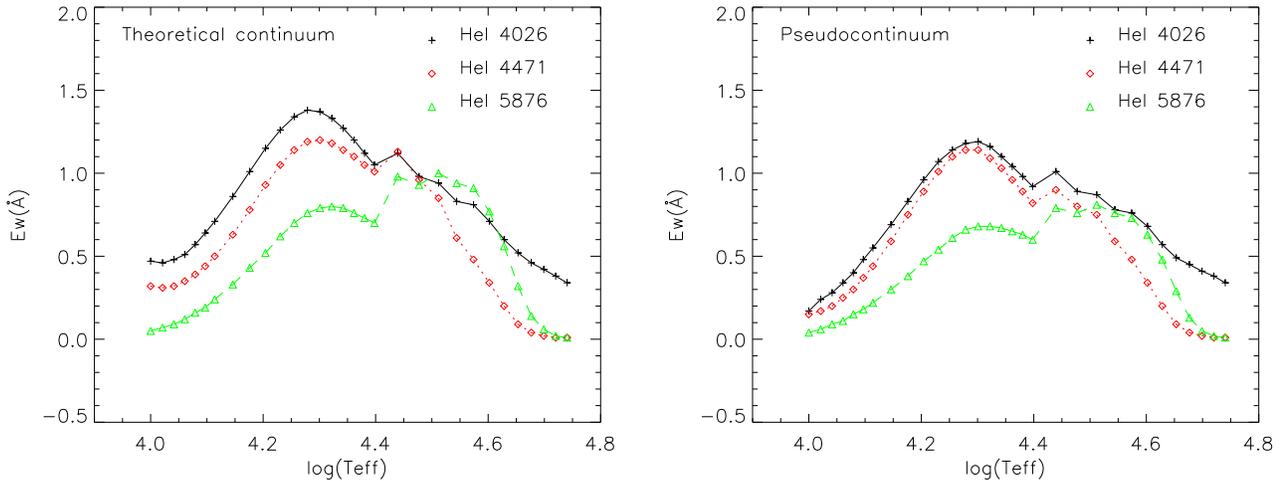}
\caption{Equivalent width of the HeI lines ($\lambda$4026, $\lambda$4471 and $\lambda$5876)
  as a function of the effective temperature for log {\it{g}} = 4.0 and solar abundance.
The left panel shows 
the equivalent width calculated as the integrated flux under
the real continuum. The right panel shows the equivalent width calculated
as the integrated flux under a pseudo-continuum defined after fitting a linear function to the continuum
windows defined in Table 2.}
\end{figure*}

\begin{figure} 
\includegraphics[scale=0.5]{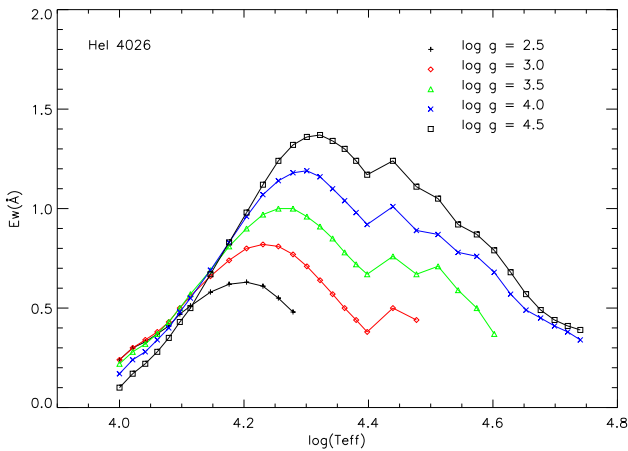}
\caption{Equivalent width of HeI $\lambda$4026 as a function of the effective temperature. Each curve corresponds to a different value of the gravity.}
\end{figure}

One of the most challenging problems in interpreting an observed galaxy spectrum is
the difficulty to disentangle age and metallicity effects. Until very recently
this analysis has relied metallic indices. Particularly effective is the 
comparison between lines which are mainly sensitive to the main sequence turnoff temperature,
like the Balmer lines, and 
metallic indices sensitive to the temperature of the 
giant branch, which depends on metallicity (Worthey 1994; Buzzoni, Mantegazza \& Gariboldi 1994).
However, it is clear that fitting the whole spectrum instead of calculating
indices should be much more efficient. This is a major motivation for generating this
library.

As a consistency check,
we calculated several widely used metal indices to test the behavior of our library. The
Balmer discontinuity is an important age diagnostic in the optical continuum of a stellar
population. While the 4000~\AA\ break is produced by a blend of metallic lines around
3800 -- 4000~\AA, the Balmer discontinuity depends on the convergence of the Balmer series
below 4000~\AA.  Different definitions can be adopted to measure this discontinuity, depending
on the stellar age we are interested to measure.  
The 4000~\AA\ break defined by Balogh (D4000, Balogh et al. 1999) is suitable to study the general
change of the discontinuity for intermediate and  old stellar populations while the Balmer discontinuity
defined by Rose (BD, Rose, Stetson, \& Tripico 1987) is used to date young stellar populations.
Balogh et al. use 100 \AA\ continuum
bandpasses to measure the break (3850 -- 3950~\AA\ and 4000 -- 4100~\AA). The Balmer
discontinuity as defined by Rose is the ratio of the average flux in two narrow
bands at 3700 -- 3825~\AA\ and 3525 -- 3600~\AA. 
Recently, the 4000~\AA\ break has been used in
combination with the H$\delta$ absorption line to date stellar populations
of galaxies in the Sloan Digital Sky Survey release (Kauffmann et al. 20003a,b,c). The definition of H$\delta_{A}$ of
Worthey \& Ottaviani (1997)
uses a central bandpass (4082 -- 4122 \AA) bracketed by two pseudo-continuum bandpasses (4030 --
4082 \AA\ and 4122 -- 4170 \AA).
Figure 15 shows D4000 as a function of the equivalent width of H$\delta$ defined by
Worthey \& Ottaviani (1997) (Ew(H$\delta_{A}$)) for different values of gravity
(left panel) and chemical abundance (right panel). Note the strong correlation between these indices
for intermediate and low temperature stars (D4000 decreases with temperature) making this plot suitable
for studying intermediate and old stellar populations.
Figure 16 shows D4000 and BD as a function of {\it{T$_{\rm{eff}}$}} for different values
of log {\it{g}} and chemical abundances.

Rose et al. (1987) proposed to use the ratio of the central line depths
of two neighboring spectral line features, which has several advantages with respect
to an equivalent width index. In particular, it is independent of the placement of the
pseudo-continuum, insensitive to reddening and only slightly dependent of the resolution.
Figure 17 is a plot of the Rose indices CaII vs. of H$\delta$/FeI for different values of
gravity and chemical abundances. Figure 18 shows each of these indices individually as a function of 
{\it{T$_{\rm{eff}}$}}
for different  values of gravity (left panels) and chemical abundance (right panels). 
CaII is the ratio of the residual central intensity of CaII + H$\epsilon$ and CaIIK lines. H$\delta$/FeI is the ratio 
of the residual central intensity of H$\delta$ to that of the FeI$\lambda$4045 line.
Figure 19 shows the comparison
between the Rose indices for different spectral resolutions of the original spectra. Our models
confirm the small dependence of these indices with resolution, up to a temperature of $\sim$~7000~K.
Below this temperature, the dependence becomes stronger. However, this relation is not useful for
such low temperatures since the H lines are very weak.

\begin{figure*}
\includegraphics[angle=90,scale=0.65]{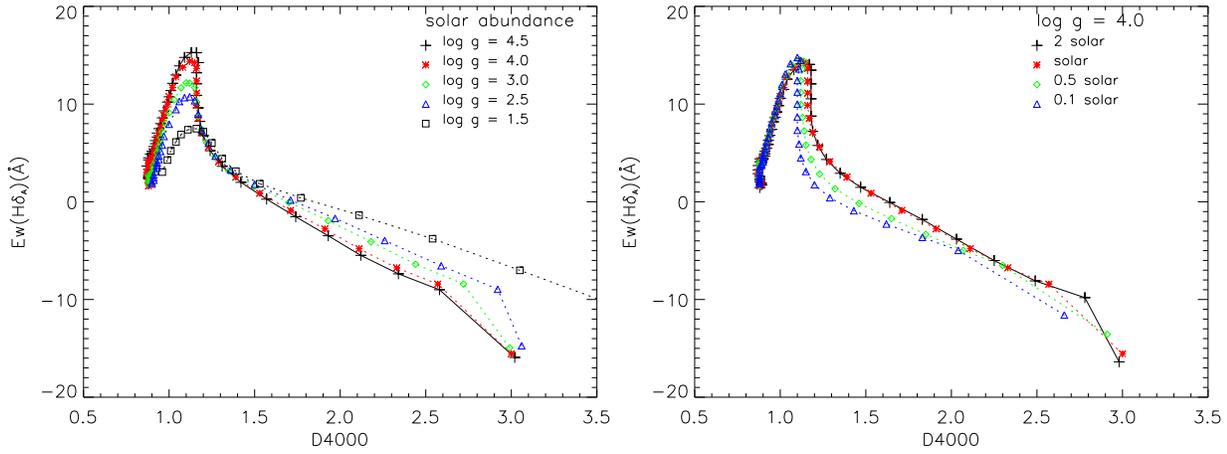}
\caption{Behavior of D4000 versus Ew(H$\delta_{A}$)
for different gravities (left) and chemical abundances (right).}
\end{figure*} 

\begin{figure*}
\includegraphics[angle=90,scale=0.65]{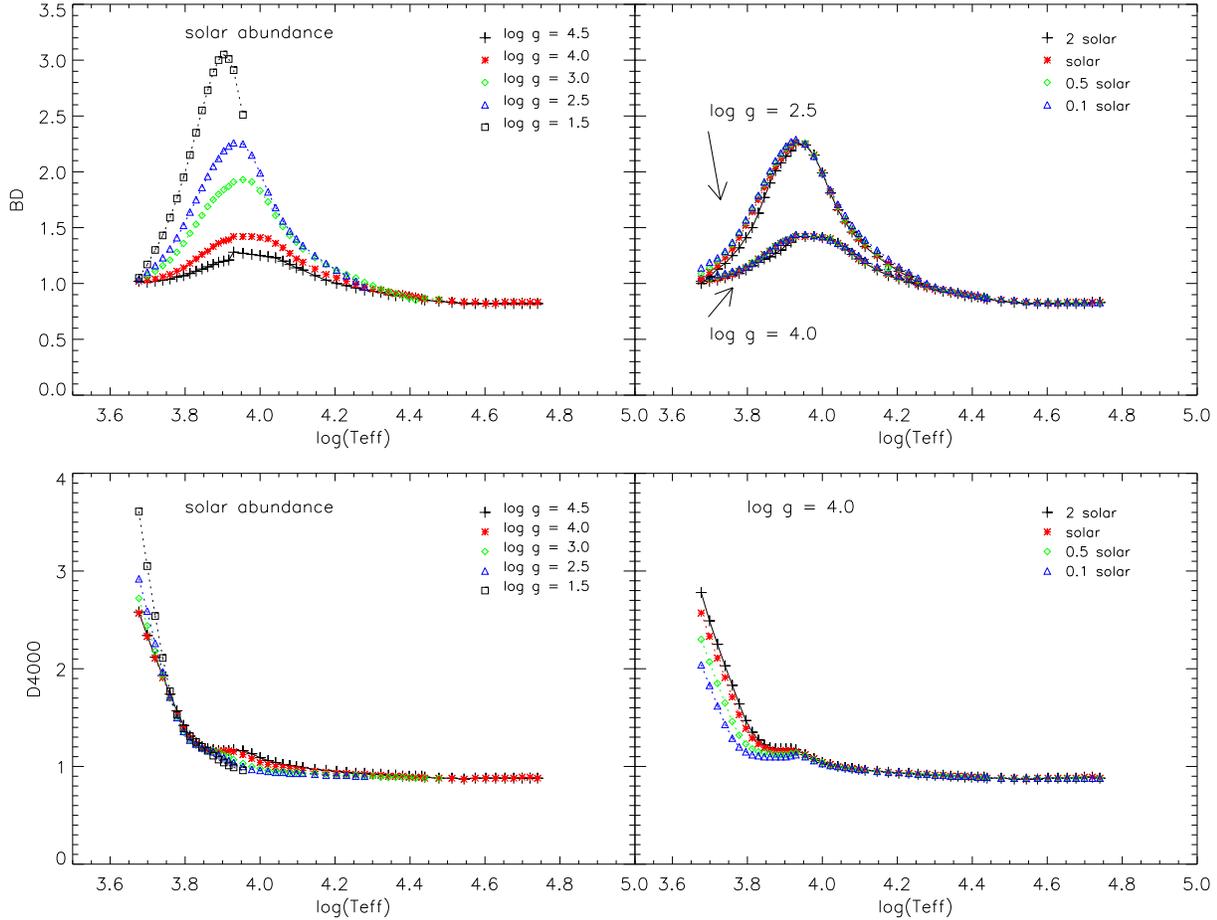}
\caption{Behavior of D4000 and BD with {\it{T$_{\rm{eff}}$}}. Top left: BD vs. {\it T$_{\rm{eff}}$}
for different log {\it g} at solar abundance; top right: BD vs. {\it T$_{\rm{eff}}$} for different
abundances and log {\it g}=2.5 and 4.5; bottom left: D4000 vs. {\it{T$_{\rm{eff}}$}} 
for different log {\it g} at solar abundance; bottom right: D4000 vs. {\it T$_{\rm{eff}}$} for different
abundances and log {\it g}=4.0.}
\end{figure*} 

 \begin{figure*}
\includegraphics[angle=90, scale=0.65]{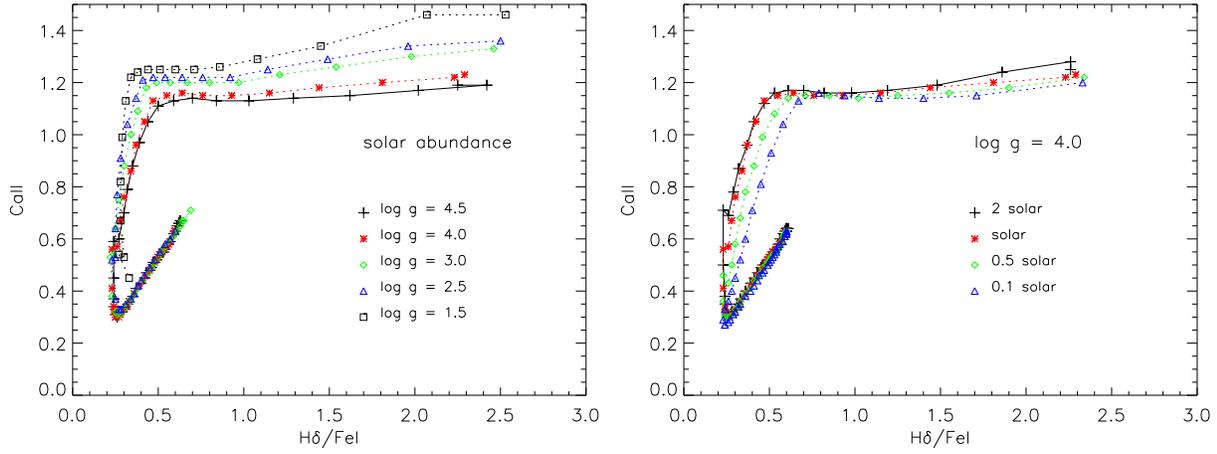}
\caption{Behavior of the Rose indices CaII and H$\delta$/FeI for different gravities (left)
and chemical abundances (right).}
\end{figure*}

 \begin{figure*}
\includegraphics[angle=90, scale=0.65]{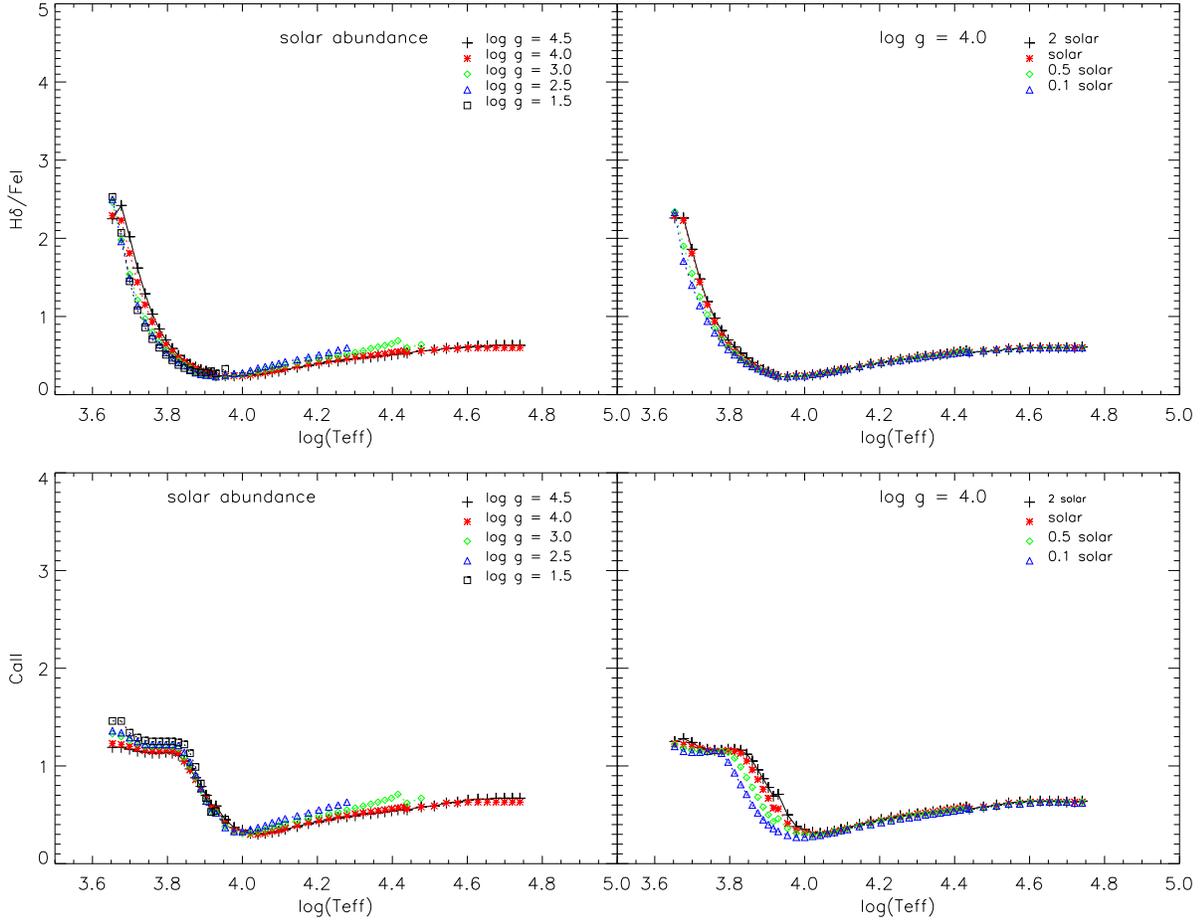}
\caption{Behavior of the Rose indices H$\delta$/FeI and CaII with {\it{T$_{\rm{eff}}$}}. Top left: H$\delta$/FeI vs. 
{\it{T$_{\rm{eff}}$}} 
for different log {\it g} at solar abundance; top right: H$\delta$/FeI vs. {\it T$_{\rm{eff}}$} for different
abundances and log {\it g}=4.0; bottom left: CaII vs. {\it{T$_{\rm{eff}}$}} 
for different log {\it g} at solar abundance; bottom right: CaII vs. {\it T$_{\rm{eff}}$} for different
abundances and log {\it g}=4.0.}
\end{figure*}

\begin{figure*}
\includegraphics[angle=90,scale=0.5]{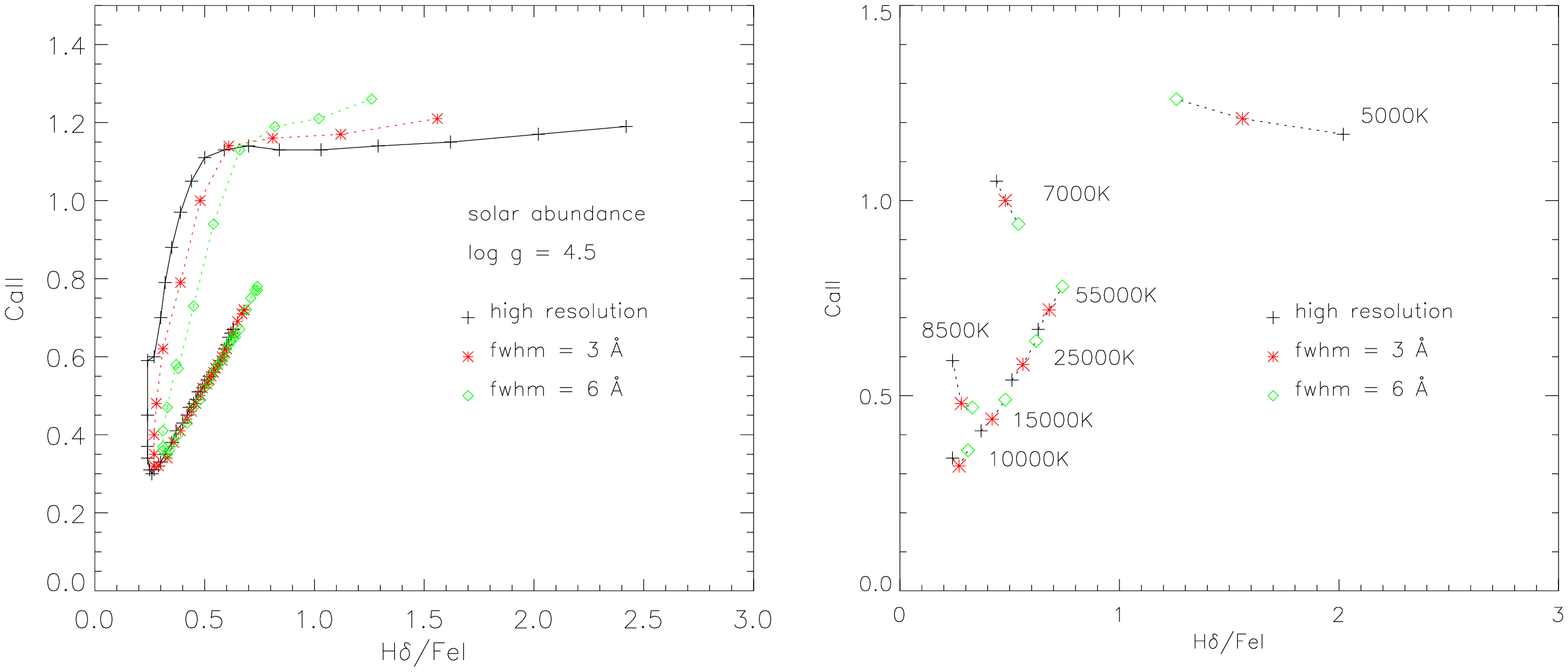}
\caption{Rose indices measured in the original spectra and 
in the spectra degraded to a resolution of fwhm= 3~\AA\ and 6~\AA\ (left).
In the right panel we restricted the data points to a few representative temperatures and connected models
with identical temperatures and different resolution to improve clarity.}
\end{figure*}

In a series of papers starting in mid 1980's, Bica \& co-workers have
explored a stellar populations synthesis technique known as Empirical
Population Synthesis, which decomposes a given galaxy spectrum 
into a combination of observed spectra of star clusters (Bica \& Alloin 1986,
Bica 1988, and Bica, Alloin, \& Schmitt 1994). In practice, instead
of modeling the {\it F}$_{\lambda}$ spectrum directly, this method synthesizes a number
of absorption line equivalent widths and continuum colors, which are used as a compact
representation of {\it F}$_{\lambda}$. The equivalent widths and colors are measured
relative to a pseudo-continuum defined at pre-selected pivot wavelengths, which is traced
interactively over the observed spectrum (see Cid Fernandes, Storchi-Bergmann,
\& Schmitt 1998 for an illustrated
discussion).
We present some indices measured in Bica's system. Figure 20 shows the dependence
of Ew(CaIIK), Ew(G~band) and Ew(MgII) on temperature for different values of chemical
abundance and gravity. The windows used for each of these indices are 3908 -- 3952~\AA,
4284 -- 4318~\AA, and 5156 -- 5196~\AA, respectively.
All of these indices are very useful tools to study intermediate and old stellar
populations, since they are very sensitive for low- and intermediate-temperature stars.

 \begin{figure*} 
\includegraphics[scale=0.9]{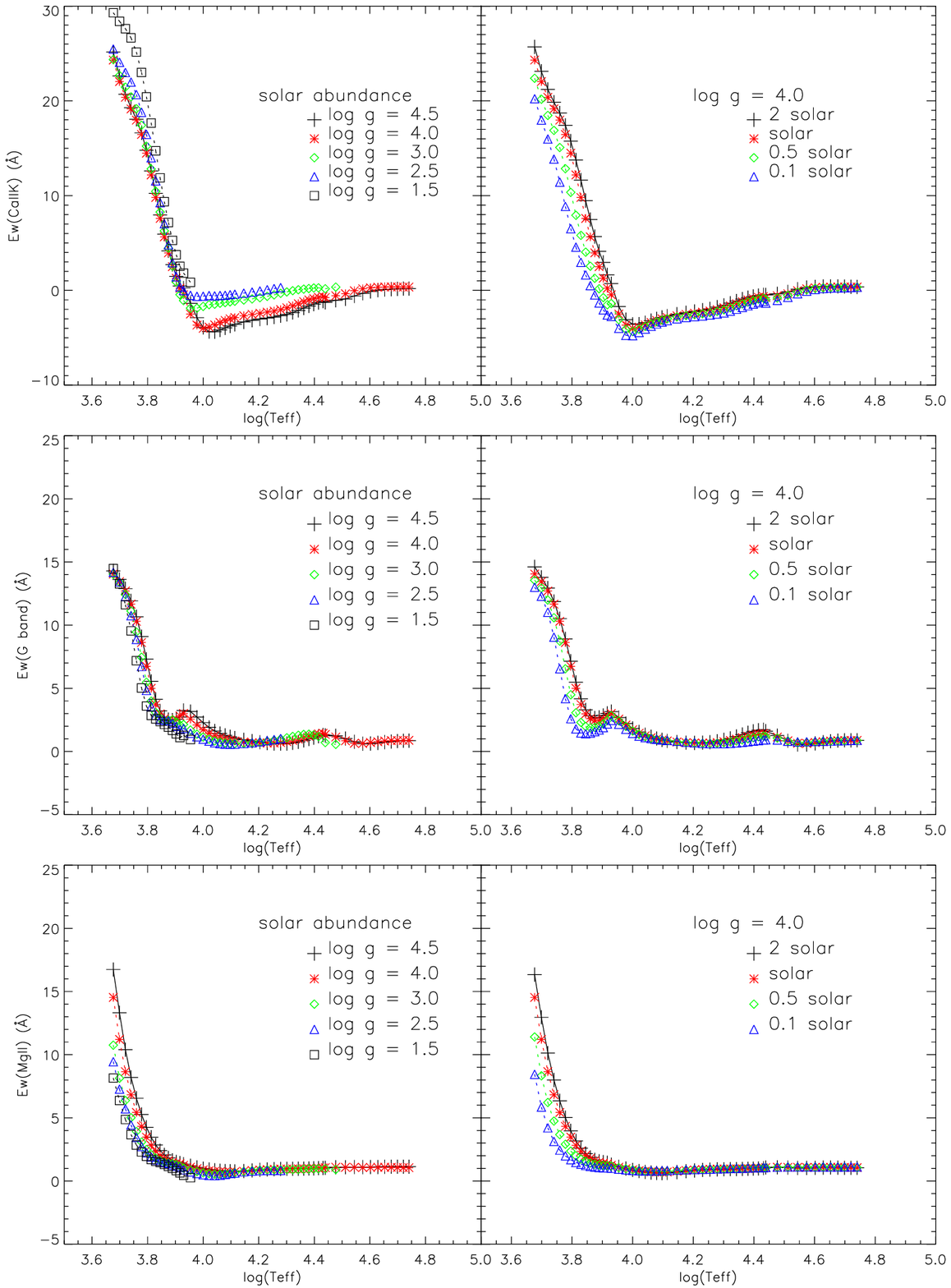}
\caption{Behavior of the Bica indices with temperature.}
\end{figure*}

\section {comparison with empirical libraries}

The decisive test for the reliability of our synthetic library spectra is a
comparison with observed stellar spectra. Recently, two empirical libraries
have  been made available by Le Borgne et al. (2003) and Valdes et al.
(2004) which are  suitable for such a test.

Le Borgne et al. (2003) compiled the STELIB library\footnote{http://webast.ast.obs-mip.fr/stelib}, 
a collection of 249
stellar  spectra at 3~\AA\ resolution covering a wide parameter space in {\it T$_{\rm{eff}}$},
log {\it g} and  [Fe/H]. We selected representative spectra whose stellar
parameters are  sufficiently close to our grid points for a useful
comparison. Since the  empirical spectra are affected by interstellar
reddening, we corrected them using  the {\it A}$_{\rm{v}}$ values given by Le Borgne et al.
The reddening law of Cardelli, Clayton,  and Mathis (1989) was used.

A comparison between our models and selected STELIB stars is in Figure 21.
The models were smoothed with a gaussian filter to the STELIB resolution of
3~\AA.  The spectra were normalized at 4020~\AA. The left panels in this figure
reproduce  the full spectra range, and the right panels show a zoomed view
of the region  around the 4000~\AA\ break. The stellar parameters given
by STELIB for each star are summarized in Table~3. For comparison, the corresponding parameters
of the synthetic spectra are  given at the top of each panel in Fig.~21. No
effort was made to interpolate in  our theoretical grid. The best fitting
model was selected from a $\chi^2$ fit of the synthetic spectra to the
observations. Considering that no optimization was performed, the agreement
over a broad parameter range is quite satisfactory. We are also reproducing
in this figure one of the worst cases: HD~48329, a G8 supergiant. Our
half-solar model appears to underestimate the blanketing around 4100~\AA. 
Remember that for this temperature range the ``half solar'' models have
actually one third solar abundances, what may be the reason for that.

In order to further explore the quality of our synthetic library at very low
temperatures, we utilized the Indo-US library of Valdes et al. (2004). This
library has particularly complete coverage at the low-temperature end.
It includes reddening corrected spectra for 1273 stars with a resolution of $\sim$~1~\AA. 
We performed a comparison with
our models following the same rules as we used for the STELIB library. The
selected stars are summarized in Table~4, and the comparison is plotted in
Figure~22. The three synthetic spectra in this temperature regime were
generated with Phoenix. Given the complexity of the line blanketing, the
agreement between models and observations is rather gratifying.

\begin{table}
\caption{Parameters of the STELIB stars used in Figure 21.}
\begin {tabular} {@{}|l|ccccc}
\hline
 Star      & Spectral type  & {\it{T$_{\rm{eff}}$}} & log {\it{g}} & [Fe/H] & {\it A}$_{\rm{v}}$  \\
\hline
HD 034816 & B0.5IV   &  30262 K  &  4.18 &  0.05&  0.02 \\
     
HD 35497  & B7III    &  13622 K  & 3.80  &  0.41& 0.03 \\
HD 32537  & F0V      &  7031 K & 4.04  &   -0.23& 0.4\\
HD 48329  & G8Ib     &  4540 K & 0.88   &  -0.04& 0.0 \\
\hline
\end{tabular}
\end{table}

 \begin{table}
\caption{Parameters of the Indo-US stars used in figure 21.}
\begin {tabular} {@{}|l|cccc}
\hline
 Star      & Spectral type  & {\it{T$_{\rm{eff}}$}} & log {\it{g}} & [Fe/H] \\
\hline
HD 112300  & M3III    & 3700 K & 1.3    & -0.16  \\     
HD 149161  & K4III    & 3910 K & 1.6   & -0.23  \\
HD 39801  &  M1       & 3540 K  & 0.0  & 0.05   \\
\hline
\end{tabular}
\end{table}

\begin{figure*} 
\includegraphics[scale=0.90]{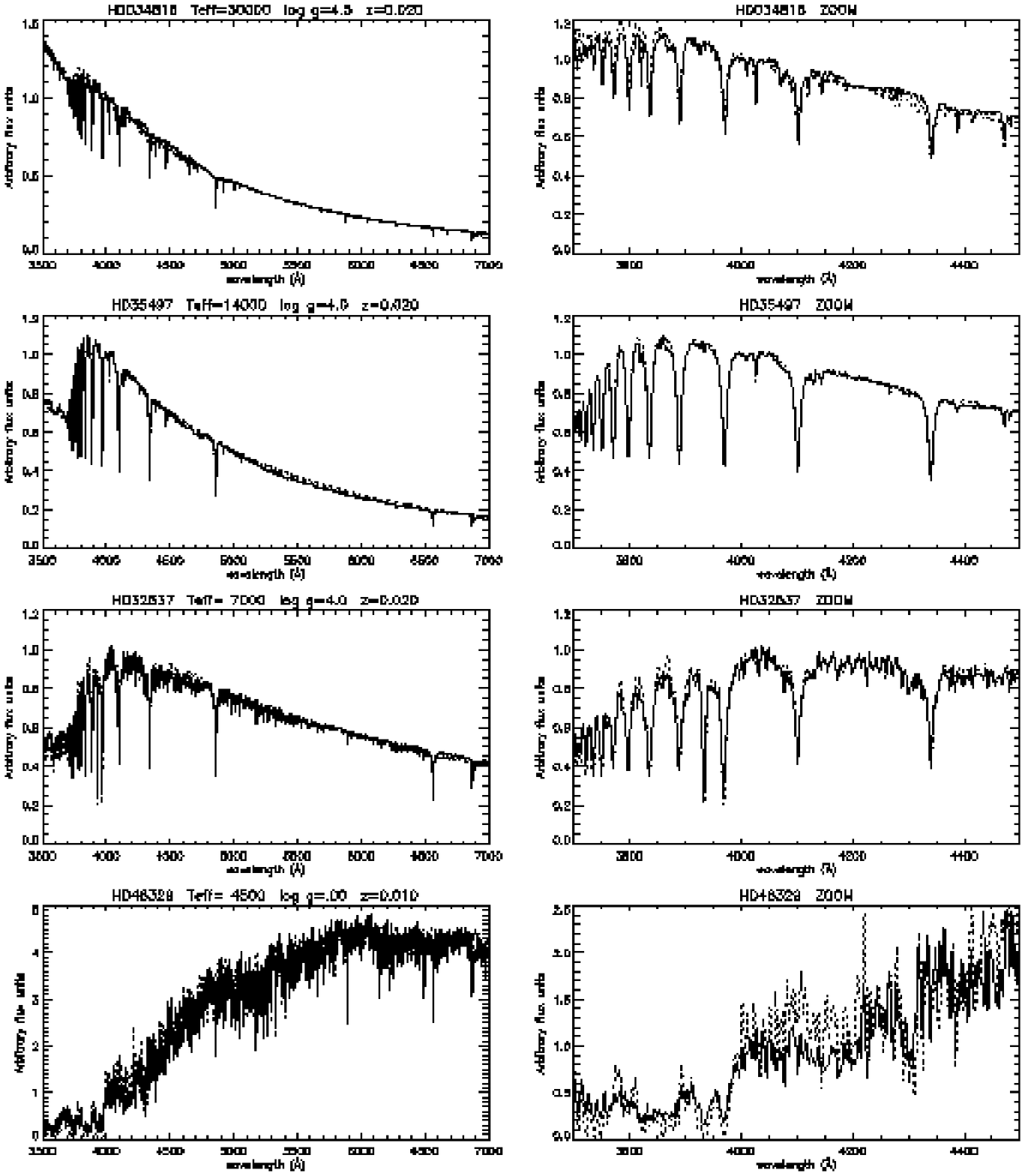}
\caption{Comparison between representative from the STELIB library (solid line) and
synthetic spectra having similar stellar parameters (dotted line). Left panels:
full wavelength range; right panels: zoomed view of the region around the 4000~\AA\ break.}
\end{figure*}

\begin{figure*} 
\includegraphics[scale=0.90]{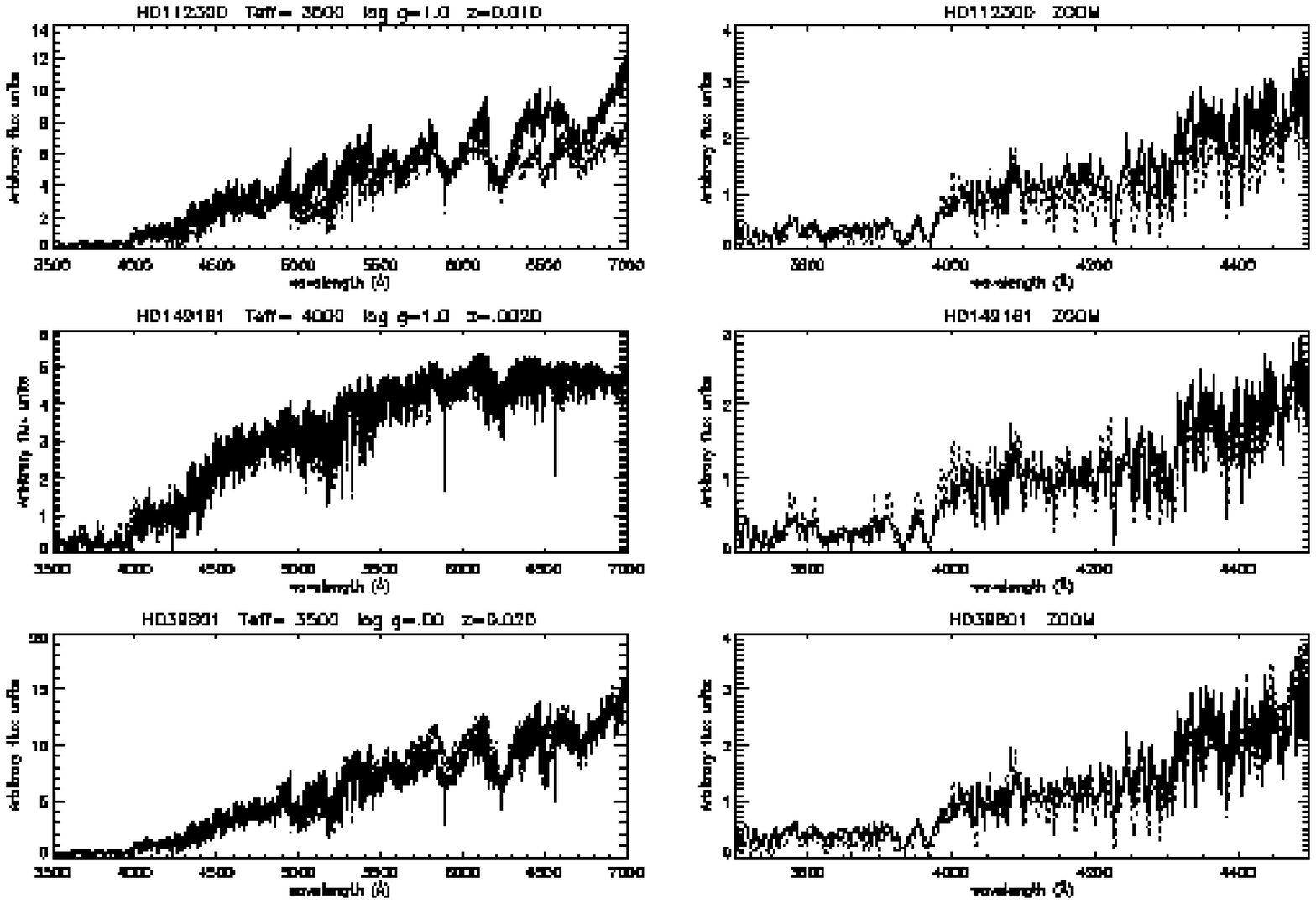}
\caption{Comparison between representative from the Indo-US library (solid line) and
synthetic spectra having similar stellar parameters (dotted line). Left panels:
full wavelength range; right panels: zoomed view of the region around the 4000~\AA\ break.}
\end{figure*}

\section{Summary and conclusions}

We generated a grid of stellar atmospheres and synthetic spectra covering the optical
range from 3000~\AA\ to 7000~\AA, with a sampling of 0.3 \AA. The grid spans a range of 
effective temperatures from 3000 K to 55000 K, and gravities from --0.5 to 5.5 for four 
different metallicity values (one tenth solar, half solar, solar and twice solar). The
spectra were generated using most up-to-date codes available in the literature for each
temperature and gravity values. The library contains 1654 spectra, incorporating the last
improvements in stellar atmospheres: non-LTE line-blanketed models for hot stars ({\it{T$_{\rm{eff}}$}} $\ge$
27500~K) and Phoenix LTE line-blanketed models for cool stars (3000~K $\leq$ {\it{T$_{\rm{eff}}$}} 
$\leq$ 4500~K).
We used Kurucz atmosphere models for the parameter space in between, but
the codes SPECTRUM and SYNSPEC were used to generate the high-resolution spectral
profiles.

We calculated equivalent widths of H and HeI lines to understand the general behavior
of our library. We found, as expected, that these lines are very sensitive to gravity and temperature,
what makes them an efficient tool for the determination of fundamental stellar parameters.
We also calculated some other metallic indices widely used in the analysis of stellar populations.
We show a comparison between 
the 4000 \AA\ break defined by Balogh et al. (1999), D4000, and the Balmer discontinuity
defined by Rose et al. (1987), BD. D4000 is suitable for studying the discontinuity for intermediate
and old stellar populations, while BD can be used to date young stellar populations. 
We also show comparisons of D4000 vs. Ew(H$\delta_A$), where Ew(H$\delta_A$) is the equivalent
width of H$\delta$ as defined by Worthey \& Ottaviani (1987). There is a strong correlation 
between these indices for low and intermediate temperature stars, making them useful
for studying intermediate and old stellar populations.
We show the behavior of the CaII and  H$\delta$/FeI indices defined by Rose et al. (1987)
and the Ew(CaIIK), Ew(G~Band) and Ew(MgII) defined by Bica (Bica \& Alloin 1986, Bica 1988) for
our library.

A final consistency check was done by comparing our theoretical models with the empirical
libraries of Le Borgne et al. (2003) and Valdes et al. (2004). We found that the agreement
between models and observations is excellent.

This library constitutes the most comprehensive theoretical stellar library for stellar population
synthesis so far, not only because of its optimal parameter space coverage, but also because
of its superior resolution
and the input physics in the models. The library is available at http://iaa.csic.es/$\sim$rosa and 
http://www.astro.iag.usp.br/$\sim$lucimara/library.html.

\section*{Acknowledgments}
We thank Roberto Terlevich and Emanuelle Bertone for very useful coments on the paper.
LPM acknowledge support of IAA (Granada). RGD and
MC acknowledge support by the Spanish Ministry of Science and Technology (MCyT) thougth grant
AYA-2001-3939-C03-01 and AYA-2001-2147-C02-01. PHH was supported in part by the P\^ole
Scientifique de Mod\'elisation Num\'erique at ENS-Lyon.

\bsp

\label{lastpage}

\end{document}